\documentclass[aps,prb,reprint,superscriptaddress,showpacs,longbibliography]{revtex4-2}
\usepackage{mathtools,amssymb,graphicx,units}
\usepackage[usenames,dvipsnames]{color}
\usepackage[plainpages=false,pdfpagelabels,colorlinks=true,linkcolor=red,urlcolor=blue,citecolor=blue,pdftitle={Title},pdfauthor={},pdfdisplaydoctitle=true,pdfduplex=DuplexFlipLongEdge]{hyperref}
\usepackage{subfigure}
\usepackage{grffile}
\usepackage{bm}
\usepackage{mhchem}
\usepackage{bibentry}
\usepackage{natbib}
\usepackage[capitalize]{cleveref}  

  

\newcommand{\abs}[1]{\ensuremath{\left| #1 \right|}}
\newcommand{\ket}[1]{\left\vert#1\right\rangle}

\newcommand{\braket}[2]{\left\langle#1\left\vert#2\right.\right\rangle}
\newcommand{\eval}[3]{\left\langle#1\left\vert#2\right\vert#3\right\rangle}
\newcommand{\vev}[1]{\left\langle #1\right\rangle}
\newcommand{\1}{\mbox{\bf 1}}
\newcommand{\ud}{\mathrm{d}}

\newcommand\redsout{\bgroup\markoverwith{\textcolor{red}{\rule[0.5ex]{2pt}{0.4pt}}}\ULon}

\begin{document}
\title{The antiferromagnetic Chern insulator phase in the Kane-Mele-Hubbard model}

\author{Bao-Qing Wang}
\affiliation{Lanzhou Center for Theoretical Physics, Key Laboratory of Quantum Theory and Applications of MoE, and Key Laboratory of Theoretical Physics of Gansu Province, Lanzhou University, Lanzhou 730000, China}

\author{Can Shao}
\affiliation{Department of Applied Physics and MIIT Key Laboratory of Semiconductor Microstructure and Quantum Sensing, Nanjing University of Science and Technology, Nanjing 210094, China}

\author{Takami Tohyama}
\affiliation{Department of Applied Physics, Tokyo University of Science, Tokyo 125-8585, Japan}

\author{Hong-Gang Luo}
\affiliation{Lanzhou Center for Theoretical Physics, Key Laboratory of Quantum Theory and Applications of MoE, and Key Laboratory of Theoretical Physics of Gansu Province, Lanzhou University, Lanzhou 730000, China}
\affiliation{Beijing Computational Science Research Center, Beijing 100084, China}

\author{Hantao Lu}
\email{luht@lzu.edu.cn}
\affiliation{Lanzhou Center for Theoretical Physics, Key Laboratory of Quantum Theory and Applications of MoE, and Key Laboratory of Theoretical Physics of Gansu Province, Lanzhou University, Lanzhou 730000, China}

\date{\today}

\begin{abstract}
The emergence of the antiferromagnetic (AFM) Chern insulator (AFCI) phase in the Kane-Mele-Hubbard (KMH) model with a finite sublattice potential is investigated. The AFCI, characterized by AFM correlations coexisting with quantized Hall conductance, has long raised the question of whether it can exist in the KMH model that respects time-reversal symmetry (TRS). Using exact diagonalization, we analyze the excitation gap, anisotropic AFM correlations along the $z$ axis and in the $xy$ plane, and the fidelity susceptibility under twisted boundary conditions, all of which provide consistent evidence for the AFCI phase. In particular, our numerical evaluation on the (spin) Chern number reveals a breakdown of adiabatic continuity in the twist-angle space, indicating an instability toward TRS breaking driven by Hubbard-induced AFM perturbations. A modified computational scheme is further proposed, which yields a robust quantized Chern number $C=1$ within this phase. 
\end{abstract}

\maketitle

\section{Introduction}
\label{sec_intr}

The interplay between topology and strong correlations has become a central theme in modern condense matter physics~\cite{Hohenadler_2013,Rachel_2018}. In particular, the emergence of interaction-driven magnetic orders, when combined with topologically nontrivial band structures, provides a fertile ground for exploring novel correlated topological phenomena. Such an interplay has led to the discovery of a variety of quantum phases with rich and unconventional topological characteristics.

A prominent example is the antiferromagnetic Chern insulator (AFCI), characterized by antiferromagnetic (AFM) correlations coexisting with quantized Hall conductance. First proposed in Ref.~\cite{He_2011} through an effective and mean-field (MF) analysis of the spinful Haldane-Hubbard (HH) model, the quantum state combines AFM order that is robust against external magnetic perturbations, with spin-selective chiral edge modes protected by topology, offering potential advantages for spintronics and quantum computation. Since its original proposal, the AFCI phase has been identified in a number of theoretical studies across distinct systems and platforms~\cite{Vanhala_2016,JiangK_2018,Wang_2018,Wang_2019,Ebrahimkhas_2021,Ebrahimkhas_2022,Torbati_2024,Torbati_2024b,Rubem_2024,Shao_2024,Wang_2024}, highlighting the universality of its formation mechanisms.

In general, the AFCI phase arises when one spin sector forms a quantum Hall (QH) state while the other remains topologically trivial, necessitating genuine time-reversal symmetry (TRS) breaking that cannot be restored by any space-group operation~\cite{JiangK_2018,Wang_2019,Ebrahimkhas_2021,Ebrahimkhas_2022}. Consequently, a finite sublattice potential $\Delta_{\mathrm{AB}}$ is typically required to stabilize the phase~\cite{Wang_2024,Torbati_2025}.

Two paradigmatic unperturbed topological band structures have been widely employed in the study of the AFCI. The first is the Haldane model on the honeycomb lattice~\cite{Haldane_1988}, where TRS is broken explicitly and both spin components share identical dispersions. The second is the Kane-Mele (KM) model in which spin-orbit coupling (SOC) generates topologically nontrivial bands while preserving TRS~\cite{Kane_2005a,Kane_2005b}. Other topologically nontrivial systems, such as the Harper-Hofstadter model, have also been considered~\cite{Ebrahimkhas_2021,Ebrahimkhas_2022,Torbati_2025,Wang_2018,Wang_2019}. On the interaction side, the onsite Hubbard $U$ is typically adopted to capture correlation effects, although the influence of longer-range interactions, such as nearest-neighbor terms, has also been explored~\cite{Wang_2019,Shao_2021,Wang_2024}.

The existence of the AFCI phase in the HH model with sublattice potential $\Delta_{\mathrm{AB}}$, where TRS is explicitly broken at the single-particle level, has been firmly established through various approaches, including MF theory, dynamical mean-field theory (DMFT), exact diagonalization (ED), diagrammatic quantum Monte Carlo analysis, and density-matrix renormalization group (DMRG) calculations~\cite{He_2011,Vanhala_2016,Tupitsyn_2019,HeWX_2024}. In contrast, its realization in the Kane-Mele-Hubbard (KMH) model that preserves TRS, remains a subject of ongoing debate. 

MF studies on the KMH model with sublattice potential suggests the appearance of the AFCI phase at large values of both $U$ and $\Delta_{\mathrm{AB}}$~\cite{JiangK_2018,XiaoDi_2024b}. Slave-boson analyses from the strong-coupling limit yield qualitatively similar phase diagrams, with the magnetic boundaries shifted to higher $U$~\cite{JiangK_2018}. In contrast, determinant quantum Monte Carlo (DQMC) simulations, where the topological response is inferred from the compressibility under applied flux, have reported the absence of a quantum Hall phase, instead finding an incipient quantum spin Hall (QSH) phase at large $U$ and $\Delta_{\mathrm{AB}}$, along with an extended quasisemimetallic region on the boundaries of the band insulator (BI) and QSH phases~\cite{Phillips_2025}. Interestingly, recent real-space DMFT studies of the Kane-Mele-Kondo (KMK) model---where interactions arise from the coupling between itinerant electrons and localized spins---have once again identified the AFCI phase~\cite{Torbati_2024,Torbati_2024b}.

Motivated by these developments and the remaining controversy regarding the KMH model, we present in this work a detailed ED study of the half-filled KMH model with finite sublattice potential, focusing on the possible realization of the AFCI phase. As a numerically exact and unbiased approach, ED provides precise information on the ground state and low-lying excitations, provided that finite-size effects are properly controlled. Previous studies demonstrated that even relatively small clusters can capture the essential topological features of interacting systems, including edge conductance and topological transition points~\cite{Varney_2010,Varney_2011,Rubem_2021}. In particular, the (spin) Chern number can be computed directly by integrating the Berry curvature over the parameter space of boundary twists, offering a robust and gauge-invariant measure of topological character~\cite{Thouless_1982}. 

In this study, we identify a continuous region in the phase diagram---denoted as the C1 phase---characterized by a notably small excitation gap. The analysis of local observables, particularly the spin-density-wave (SDW) orders along the $z$ axis and in the $xy$ plane, reveals distinct behavior in this region compared with neighboring phases. Furthermore, the fidelity susceptibility with respect to twisted boundary conditions (TBC) indicates that the C1 phase possesses a distinct topological character. The breakdown of adiabatic continuity in the twist-angle space signals a tendency toward TRS breaking driven by correlations. By introducing a modified numerical procedure for evaluating the Chern number, we find $C=1$ within the C1 region, providing direct evidence for the AFCI phase in the KMH model. 

The remainder of this paper is organized as follows. In Sec.~\ref{sec_model} we introduce the KMH model and define local SDW and charge-density-wave (CDW) orders. The numerical framework for evaluating topological invariants within ED is also outlined. In Sec.~\ref{sec_results_ED}, which consists of five subsections, we present the ED results and analyze the corresponding phase diagram in detail. Section~\ref{sec_results_MF} provides an intuitive MF interpretation of the topological phase transitions in terms of an effective sublattice potential. Finally, we conclude our paper in Sec.~\ref{sec_summary}. 

\section{Model and observables}
\label{sec_model}

In this article, we study the KMH model~\cite{Rachel_2010} at half filling in the presence of a staggered sublattice potential. By incorporating the onsite Hubbard interaction into the original Kane-Mele model~\cite{Kane_2005a,Kane_2005b}, the KMH model provides a paradigmatic framework for exploring the interplay between SOC-induced topological band structures and electron-electron interactions in two-dimensional (2D) lattices. 

\subsection{Model}

The Hamiltonian of the KMH model reads
\begin{equation}
\hat{H}=\hat{H}_{0}+\hat{H}_{\mathrm{I}},
\label{eq_ham}
\end{equation}
with the noninteracting part~\cite{Kane_2005b} given by
\begin{eqnarray}
\hat{H}_{0}&=&-t_{1}\sum\limits_{\vev{ij},\sigma}{c}_{i\sigma}^{\dag}{c}_{j\sigma}-i t_{2}\sum\limits_{\ll ij\gg,\sigma}\nu_{ij}{c}_{i\sigma}^{\dag}\sigma_{z}{c}_{j\sigma} \notag \\
&&+\Delta_{\mathrm{AB}}\sum\limits_{i,\sigma}\xi_i\hat{n}_{i\sigma},
\label{eq_ham_0}
\end{eqnarray}
and the repulsive Hubbard-interaction part ($U>0$)
\begin{equation}
\hat{H}_{\mathrm{I}}=
U\sum\limits_{i}
\hat{n}_{i\uparrow}\hat{n}_{i\downarrow}.
\label{eq_ham_I}
\end{equation}
Here, ${c}_{i\sigma}^{\dagger}$ (${c}_{i\sigma}$) denotes the creation (annihilation) operator for an electron with spin $\sigma$ ($\sigma=\uparrow,\,\downarrow$) at site $i$, and $\hat{n}_{i\sigma}={c}^{\dagger}_{i\sigma}{c}_{i\sigma}$ is the corresponding number operator. The nearest-neighbor (NN) and the next-nearest-neighbor (NNN) hopping amplitudes are denoted by $t_1$ and $t_2$, respectively, while the staggered sublattice potential is given by $\Delta_{\mathrm{AB}}$, with $\xi_i=+1\ (-1)$ for sites $i$ on sublattice $A$ ($B$). The factor $\nu_{ij}=\pm 1$ in the intrinsic SOC term (associated with $t_2$) encodes the chirality of the hopping path from site $j$ to its NNN site $i$: say, a left turn yields $+1$ while a right turn yields $-1$~\cite{Haldane_1988}. Throughout the paper, we focus on the ground-state phase diagram of the KMH model at half filling. We set the reduced Planck constant $\hbar=1$ and $t_1$ as the unit of energy. The NNN hopping amplitude $t_2$ is fixed at $0.2$. 

In the noninteracting part of the Hamiltonian $\hat{H}_0$, TRS is preserved by incorporating the Pauli matrix $\sigma_z$ into the intrinsic SOC term. This inclusion assigns opposite signs to the hopping amplitudes for the two spin species, ensuring a total Chern number of zero. The absence of the Rashba coupling guarantees the conservation of the spin component $S^z$, allowing $\hat{H}_0$ to decompose into two decoupled Haldane models for spin-up and spin-down electrons~\cite{Haldane_1988}. Each spin sector then carries a quantized Chern number---for instance, $C_\uparrow = +1$ and $C_\downarrow = -1$ in the absence of $\Delta_{\mathrm{AB}}$---yielding a net spin Chern number $C_s = \frac{1}{2}(C_\uparrow - C_\downarrow) = 1$~\cite{ShengDN_2006,Prodan_2009}. This spin Chern number characterizes the QSH insulating phase realized in this regime. Conversely, the sublattice potential $\Delta_{\mathrm{AB}}$ explicitly breaks inversion symmetry, and in the absence of interaction drives the system into a topologically trivial, charge-ordered band insulating (BI) state when $\Delta_{\mathrm{AB}} > \Delta_{\mathrm{AB}}^c = 3\sqrt{3} t_2$. 

Turning on the Hubbard interactions in $\hat{H}_{\mathrm{I}}$ introduces electron-electron correlations into the system, enriching the phase diagram with a variety of competing phases (see, e.g., Fig.~\ref{fig_phase_ED}). In the large-$U$ limit, the system is naturally expected to enter a topologically trivial phase with easy-plane AFM order~\cite{Rachel_2010}. However, owing to the sensitivity of the QSH-BI phase boundary to Hubbard-induced AFM perturbations, it is predicted by the mean-field analysis that the AFCI phase with Chern number $C=1$ emerges~\cite{JiangK_2018,XiaoDi_2024b}. In this phase, TRS is spontaneously broken, and the AFM order reorients from the easy-plane to the easy-axis direction. The present work is devoted to a detailed study of this AFCI phase. 

\subsection{Local orders}

To characterize various phases, both local order parameters---namely the SDW and CDW orders---and the topological invariants, including the Chern number and the spin Chern number, are evaluated in the study. 

The SDW and CDW orders dominate in the large-$U$ and large-$\Delta_{\mathrm{AB}}$ limits, respectively. In the ED calculation, the orders are quantified via the staggered structure factors, defined as follows~\cite{Shao_2021,Wang_2024,Phillips_2025}:
\begin{subequations}  \label{eq_SDWCDW_ED}
\begin{align}
S_{\mathrm{AFM}}^{zz}&=\frac{1}{N}\sum_{i,j}(-1)^{\eta}\vev{\hat{S}_i^z\hat{S}_j^z} \label{eq_SDWzz_ED},\\
S_{\mathrm{AFM}}^{xy}&
=\frac{1}{4N}\sum_{i,j}(-1)^{\eta}\vev{\hat{S}_i^+\hat{S}_j^-+\hat{S}_i^-\hat{S}_j^+},\label{eq_SDWxy_ED}\\
S_{\mathrm{CDW}}&=\frac{1}{N}\sum_{i,j}(-1)^{\eta}\vev{\left(\hat{n}_{i\uparrow}+\hat{n}_{i\downarrow}\right)\left(\hat{n}_{j\uparrow}+\hat{n}_{j\downarrow}\right)}.
\label{eq_CDW_ED}
\end{align}
\end{subequations}
In the above equations, $\eta = 0$ ($\eta = 1$) if sites $i$ and $j$ belong to the same (different) sublattice. The spin operators read $\hat{S}_i^z=\left(\hat{n}_{i\uparrow}-\hat{n}_{i\downarrow}\right)/2$, $\hat{S}_i^+=c_{i\uparrow}^{\dagger}c_{i\downarrow}$, and so on. From these expressions, it is evident that $S_{\mathrm{AFM}}$ and $S_{\text{CDW}}$ characterize the AFM correlations and the charge modulations between the $A$ and $B$ sublattices, respectively. The previously mentioned BI phase exhibiting charge order is henceforth referred to as the CDW phase. In contrast to the HH model, it is essential in the KMH model to distinguish AFM correlations along the $z$ axis from those in the $xy$-plane since the intrinsic SOC [Eq.~\eqref{eq_ham_0}] already reduces the spin symmetry from $\mathrm{SU}(2)$ to $\mathrm{U}(1)$.

In contrast, within the MF calculation, the (overall) SDW and CDW order parameters can be written as
\begin{align}
\mathcal{O}_{\text{SDW}}&=\abs{\frac{1}{2}\left(\vev{\vec{S}_{\mathrm{A}}}-\vev{\vec{S}_{\mathrm{B}}}\right)},\notag \\
\mathcal{O}_{\text{CDW}}&=\abs{\left(n^{\mathrm{A}}_{\uparrow}+n^{\mathrm{A}}_{\downarrow}\right)-\left(n^{\mathrm{B}}_{\uparrow}+n^{\mathrm{B}}_{\downarrow}\right)}.
\label{eq_SDWCDW_MF}
\end{align}
Here, $\vec{S}_{\mathrm{A}(\mathrm{B})}$ denotes the MF magnetization vector on the $A$($B$) sublattice, and $n^{\alpha}_{\sigma}$ represents the MF expectation value of the spin-$\sigma$ electron density on sublattice $\alpha$.

\subsection{Topological invariants}

The classification of topological phases in quantum systems fundamentally relies on integer-valued invariants that remain robust against adiabatic and continuous deformations of the Hamiltonians~\cite{Hasan_2010,ZhangSC_2011,Chiu_2016}. Central to the present study are the Chern number~\cite{Thouless_1982} and spin Chern number~\cite{ShengDN_2006,Fukui_2007}, which serve as the topological invariants characterizing the QH and QSH effects, respectively. These invariants are directly associated with the presence of symmetry-protected edge states and quantized transport responses. 

In noninteracting systems with translational symmetry, the Chern number of a given Bloch band with eigenstates $\ket{n(\mathbf{k})}$ is computed by integrating the Berry curvature over the Brillouin zone~\cite{Thouless_1982}
\begin{equation}
C_n = \frac{1}{2\pi} \int_{\text{BZ}}   {\ud}^2k  \, F(\mathbf{k}),
\label{eq_cn}
\end{equation}
where the Berry curvature $F(\mathbf{k}) = \partial_x A_y(\mathbf{k}) - \partial_y A_x(\mathbf{k})$, and the connection $A_\mu(\mathbf{k}) = -i\eval{n(\mathbf{k})}{\partial_\mu}{n(\mathbf{k})}$ for the band. Here, $\partial_{\mu} \equiv \partial / \partial k_{\mu}$ with $\mu = x, y$. 

For a noninteracting system that preserves TRS, the total Chern number of the occupied bands, $C=\sum_n C_n$, must vanish~\cite{Moore_2007}. In such cases, the topological character of the band structure is classified by a $\mathbb{Z}_2$ invariant~\cite{Kane_2005a,ShengDN_2006,Fu_2006}. However, if a system features decoupled spin channels with conserved $S^z$---as in the KM model without Rashba SOC---a spin Chern number can be defined alternatively as~\cite{ShengDN_2006}
\begin{equation}
C_s = \frac{1}{2} \left( C_{\uparrow} - C_{\downarrow} \right),
\label{eq_cs}
\end{equation}
where $C_{\sigma}\ (\sigma=\uparrow,\downarrow)$ denotes the Chern number for spin-$\sigma$ channels. The corresponding $\mathbb{Z}_2$ invariant simply reads $\nu = C_s \bmod 2$~\cite{Hasan_2010}.

When interactions are introduced, as in the KMH model considered here, the Chern number of a many-body state can be evaluated by integrating the Berry curvature over a 2D parameter space $(\theta_x, \theta_y)$, rather than over the Brillouin zone~\cite{Niu_1985,Arovas_1988,Bradlyn_2025}
\begin{equation}
C=\iint\frac{\ud\theta_{x}\,\ud\theta_{y}}{2\pi i}
\left[\braket{\frac{\partial\Psi}{\partial{\theta_{x}}}}{\frac{\partial\Psi}{\partial\theta_{y}}}
-\braket{\frac{\partial\Psi}{\partial\theta_{y}}}{\frac{\partial\Psi}{\partial\theta_{x}}}\right].
\label{eq_Chern_theta}
\end{equation}
Here, $\theta_x$ ($\theta_y$) $\in[0,2\pi)$ denotes the twist angle along the $x$ ($y$) direction in the TBC imposed on a finite lattice
\begin{equation}
c_{j+L_x\hat{\mathbf{x}}} = e^{i\theta_x} c_j,\qquad c_{j+L_y\hat{\mathbf{y}}} = e^{i\theta_y} c_j,
\label{eq_TBC_C}
\end{equation}
where $j$ is a composite index labeling the lattice site in two dimensions, and the system size is assumed to be $L_x\times L_y$. These boundary conditions effectively place the system on a torus, with $\theta_x$ and $\theta_y$ corresponding to U(1) fluxes threaded through its noncontractible loops~\cite{Laughlin_1981}. They serve as external gauge fields coupling to the many-body wavefunction, and the resulting parameter space $(\theta_x,\theta_y)$ (denoted as $\bm{\theta}$ space) plays the role of a generalized Brillouin zone for defining Berry curvature and topological invariants in interacting systems. 

Similarly, to compute the spin Chern number, one imposes a modified TBC, with spin-dependent twisting along the $\hat{\mathbf{x}}$ direction and spin-independent twisting along the $\hat{\mathbf{y}}$ direction as~\cite{ShengDN_2006}
\begin{equation}
c_{j+L_x\hat{\mathbf{x}}} = e^{i\theta_x\sigma_z} c_j,\qquad c_{j+L_y\hat{\mathbf{y}}} = e^{i\theta_y} c_j.
\label{eq_TBC_Cs}
\end{equation}
This ensures that spin-$\uparrow$ and spin-$\downarrow$ electrons experience opposite fluxes along the $\hat{\mathbf{y}}$ direction.  

The numerical evaluation of the integration in Eq.~\eqref{eq_Chern_theta} can be efficiently performed in a discrete and gauge-invariant manner~\cite{Fukui_2005}. In our simulations, the eigenstate $\Psi(\bm{\theta})$---typically the ground state---of the Hamiltonian $\hat{H}(\bm{\theta})$ is obtained on a discrete grid in $\bm{\theta}$ space, with $\bm{\theta}=\left(\theta_x,\theta_y\right)$, via the ED method. A lattice gauge field can then be constructed accordingly. The $\mathrm{U}(1)$ link variable is defined as 
\begin{equation}
U_\mu(\theta_{\ell}) = \braket{\Psi\left(\theta_{\ell}\right)}{\Psi\left(\theta_{\ell}+\delta\theta_{\mu}\right)}/\mathcal{N}_{\mu}\left(\theta_{\ell}\right),
\label{eq_link}
\end{equation}
where $\theta_{\ell}$ denotes the grid point indexed by $\ell$ and $\delta\theta_{\mu}$ is the vector increment in the $\hat{\mu}$ direction of the $\bm{\theta}$-grid. The normalization factor is given by $\mathcal{N}_{\mu}\left(\theta_{\ell}\right)=\abs{\braket{\Psi\left(\theta_{\ell}\right)}{\Psi\left(\theta_{\ell}+\delta\theta_{\mu}\right)}}$. The lattice field strength $\tilde{F}$ is defined via the oriented product of link variables around a plaquette 
\begin{equation}
\tilde{F}\left(\theta_{\ell}\right) = \arg \left[ U_x\left(\theta_{\ell}\right) U_y\left(\theta_{\ell}+\delta\theta_x\right) U_x^{-1}\left(\theta_{\ell}+\delta\theta_y\right) U_y^{-1}\left(\theta_{\ell}\right) \right].
\label{eq_Ftilde}
\end{equation}

Finally, the (spin) Chern number---depending on whether the TBC in Eq.~\eqref{eq_TBC_C} or Eq.~\eqref{eq_TBC_Cs} is applied---can be estimated as
\begin{equation}
C = \frac{1}{2\pi} \sum_{\ell} \tilde{F}\left(\theta_{\ell}\right),
\label{eq_C}
\end{equation}
provided that the admissibility condition $\abs{F\left(\theta_{\ell}\right)}\delta\theta_x\delta\theta_y\approx\abs{\tilde{F}\left(\theta_{\ell}\right)} < \pi$ is satisfied across all plaquettes, where $F\left(\theta_{\ell}\right)$ is the Berry curvature evaluated at $\theta_{\ell}$ in the continuous limit~\cite{Fukui_2005}. The condition can, in principle, always be fulfilled with a sufficiently fine discretization, provided the {\em gap-opening condition} holds and $F\left(\theta_{\ell}\right)$ is finite everywhere---namely, that no level (band) crossing occurs throughout $\bm{\theta}$ space for the target state $\Psi(\bm{\theta})$ under continuous deformation of the Hamiltonian $H(\bm{\theta})$. In the following discussion, for clarity the lattice field strength is denoted as $\tilde{F}$ when the spin-independent TBC~\eqref{eq_TBC_C} is imposed, and as $\tilde{F}_s$ when the spin-dependent TBC~\eqref{eq_TBC_Cs} is imposed. 

\section{Exact Diagonalization Results}
\label{sec_results_ED}

\subsection{Phase diagram}

In this section, we use exact diagonalization to study the ground-state phase diagram of the KMH model~(\ref{eq_ham}) on the honeycomb lattice at half filling as a function of the Hubbard interaction $U$ and the staggered potential $\Delta_{\mathrm{AB}}$. 

\begin{figure}
\centering
\includegraphics[width=0.45\textwidth]{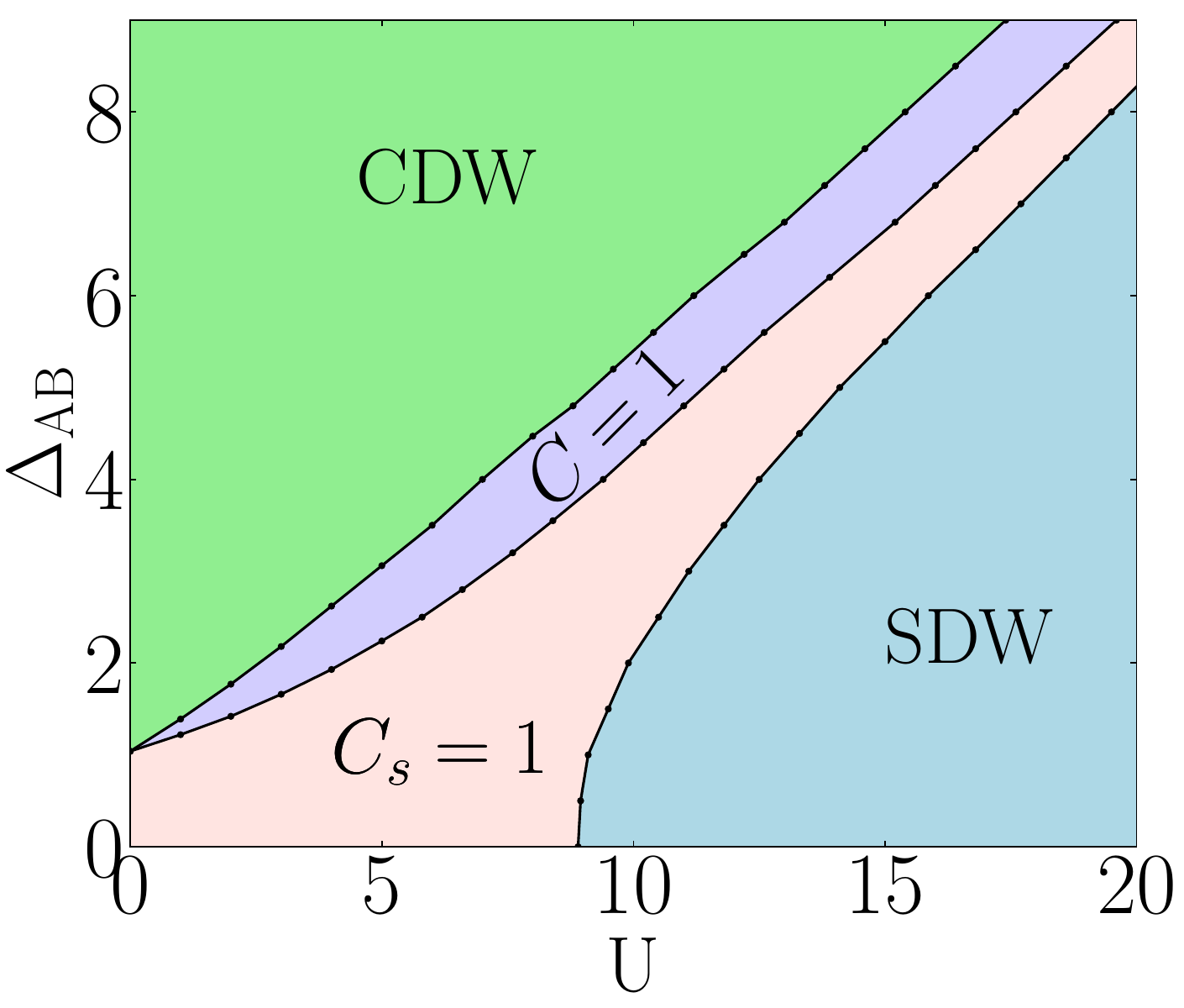}
\caption{ED Phase diagram of the KMH model~\eqref{eq_ham} in the ($U$, $\Delta_{\mathrm{AB}}$) plane with fixed $t_2=0.2$.}
\label{fig_phase_ED}
\end{figure}


Our ED calculations are performed on the so-called 12A cluster with periodic boundary condition (PBC)~\cite{Shao_2021,Varney_2010}, and the resulting phase diagram is presented in Fig.~\ref{fig_phase_ED}. For the honeycomb lattice, it has been recognized that mitigating finite-size effect and capturing essential features of the low-lying spectrum require selecting clusters whose reciprocal lattice contains the $K$ points and other high-symmetry momenta~\cite{Varney_2010,Varney_2011,Shao_2023}. This is particular important because, in the noninteracting limit, the topological phase transition involves a band inversion at the $K/K'$ points. Among all clusters up to size $18$, the 12A cluster is notable for its favorable symmetry properties---its reciprocal lattice includes the $\Gamma$ point, all the $K$ points, and a pair of $M$ points. (The detailed geometry of the 12A cluster and its $k$-point distribution can be found in Fig. 1 of Ref.~\cite{Shao_2021}.) A cluster with more complete symmetry, such as one with $24$ sites, would be preferable; however, it lies beyond our current computational capacity.

In the phase diagram shown in Fig.~\ref{fig_phase_ED}, four distinct phases are identified. As expected, two topologically trivial phases emerge in the large-$\Delta_{\mathrm{AB}}$ and large-$U$ regimes: a CDW phase characterized by charge ordering, and a SDW phase exhibiting AFM order. Supporting evidence for these orders is provided in Fig.~\ref{fig_gap-sdw}(b) (showing $S_{\mathrm{SDW}}^{zz}$) and Fig.~\ref{fig_gap4.0}(a) (showing various orders at fixed $\Delta_{\mathrm{AB}}=4.0$). In contrast, the QSH phase with $C_s=1$, which is stabilized in the small $\Delta_{\mathrm{AB}}$ and $U$ area, persists into the regime where both strengths are large, approximately following the line $U\sim 2\Delta_{\mathrm{AB}}$. This observation is consistent with those obtained from the DQMC simulations~\cite{Phillips_2025}, indicating that the competition between $U$ and $\Delta_{\mathrm{AB}}$ leaves room for the QSH phase to persist deep into the strong-coupling regime.

A prominent feature of the phase diagram in Fig.~\ref{fig_phase_ED} is the emergence of a fourth phase with Chern number $C=1$. Denoted as C1 phase here and identified in later discussions as the AFCI phase, it extends along the CDW-QSH phase boundary from the weak- to strong-coupling regime. In this phase, associated with the spin symmetry breaking that lifts the spin degeneracy, a spontaneous TRS breaking takes place and separates the spin sectors by their topological character, yielding a nonzero Chern number $C=1$~\cite{JiangK_2018}. While the existence of the AFCI phase in the HH model has been well established through various approaches~\cite{He_2011,Vanhala_2016,Tupitsyn_2019,HeWX_2024}, its realization in the KMH model remains an open issue. Although both Hartree-Fock MF and slave-particle analyses suggest its presence~\cite{JiangK_2018,XiaoDi_2024b}, recent DQMC results instead indicate a semimetallic region around the transition line between the CDW (BI) and QSH phases~\cite{Phillips_2025}. In the following Secs.~\ref{subsec_C1} and \ref{subsec_Chern}, we focus on this phase identified in our ED calculations and provide supporting evidence for its AFCI characters.

Before closing this subsection, we briefly comment on several notable features of the AFCI phase in the phase diagram. One observation is that the AFCI phase emerges even at small $U$, provided $\Delta_{\mathrm{AB}}$ is finite (e.g., it starts at $\Delta_{\mathrm{AB}}\sim 1$ as $U\to 0$ in Fig.~\ref{fig_phase_ED}), in contrast to the MF result~\cite{JiangK_2018,XiaoDi_2024b} (see also the MF phase diagram in Fig.~\ref{fig_phase_MF}). This extension of the AFCI phase into the weak-coupling regime closely resembles the behavior seen in the HH model, where it was corroborated by large-scale infinite-DMRG (iDMRG) studies~\cite{HeWX_2024}. This suggests a strong susceptibility of the CDW-QSH phase boundary to the Hubbard-induced AFM perturbations, leading to the stabilization of the AFCI phase between the two, even outside the strong-coupling regime. 

Another notable feature is the absence of a fork-like structure for the AFCI phase in the KMH model. In contrast, ED results for the HH model exhibit a bifurcated AFCI region, with one branch appearing at small $U$ and finite $\Delta_{\mathrm{AB}}$ (around $\Delta_{\mathrm{AB}}\sim 1$ when $U\to 0$, with the identical $t_2=0.2$), and another at small $\Delta_{\mathrm{AB}}$ and finite $U$ (around $U\sim 8$ as $\Delta_{\mathrm{AB}}\to 0$)~\cite{Vanhala_2016,Shao_2023}. We believe that the latter branch is likely a finite-size effect, as it disappears in iDMRG calculation~\cite{HeWX_2024}. In our ED results on KMH, such bifurcation is absent. Instead, we observe a direct phase transition from the QSH phase to the SDW phase with increasing $U$. One noticeable feature is that despite the occurrence of the topological transition, the excitation gap remains sizable at the QSH/SDW phase boundaries [Figs.~\ref{fig_gap-sdw}(a) and \ref{fig_gap4.0}(a)]. We will address this issue further in Sec.~\ref{subsec_gapclosure}.

\subsection{C1 phase} 
\label{subsec_C1}

In this subsection, we focus on the $C=1$ (C1) phase in the phase diagram (Fig.~\ref{fig_phase_ED}), analyzing its properties from several perspectives, including the excitation gap, AFM and CDW correlations, and the fidelity susceptibility with respect to TBC.

\begin{figure}
\centering
\includegraphics[width=0.45\textwidth]{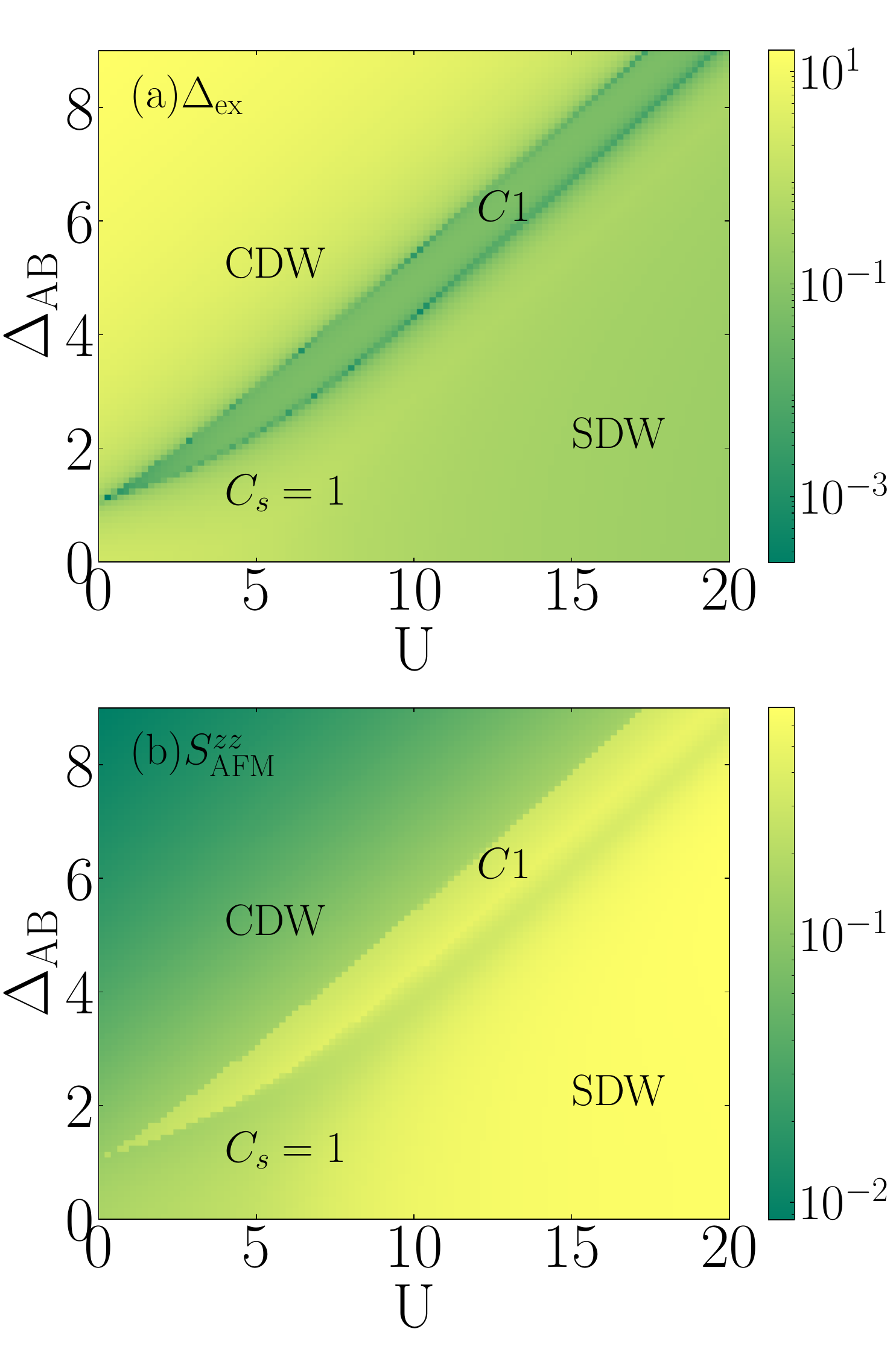}
\caption{Contour plots of (a) the excitation gap $\Delta_{\mathrm{ex}}$ and (b) the $z$-component of the AFM structure factor $S_{\mathrm{AFM}}^{zz}$, shown as functions of the Hubbard interaction $U$ and the sublattice potential $\Delta_{\mathrm{AB}}$.}
\label{fig_gap-sdw}
\end{figure}

\begin{figure}
\centering
\includegraphics[width=0.5\textwidth]{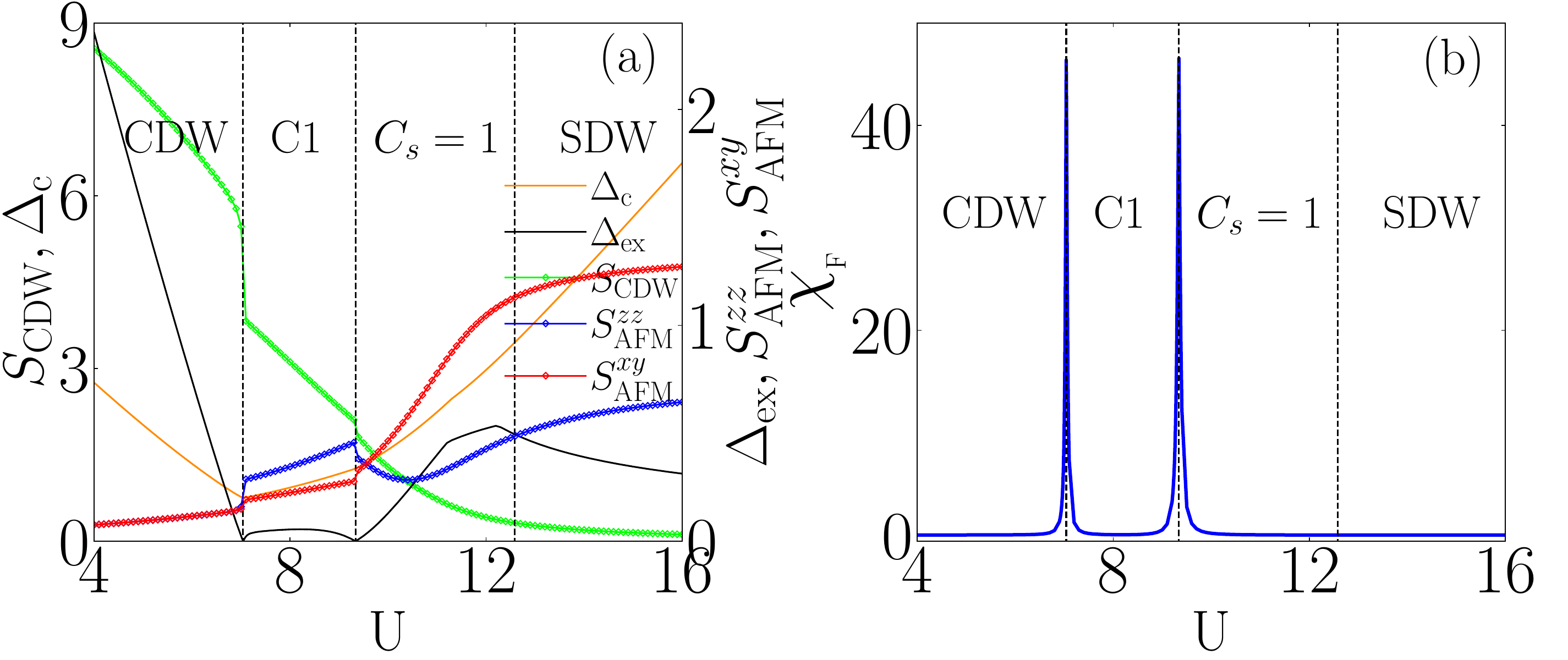}
\caption{(a) CDW and SDW structure factors, $S_{\mathrm{CDW}}$, $S_{\mathrm{AFM}}^{zz}$, and $S_{\mathrm{AFM}}^{xy}$, along with the excitation gap $\Delta_{\mathrm{ex}}$ and the single-particle charge gap $\Delta_{\mathrm{c}}$. (b) Fidelity susceptibility $\chi_F$ with respect to twisted boundary conditions. All quantities are plotted as functions of $U$ at fixed $\Delta_{\mathrm{AB}} = 4.0$. The dashed vertical lines mark the phase boundaries determined from the topological invariant calculations.}
\label{fig_gap4.0}
\end{figure}

We begin with the excitation gap $\Delta_{\mathrm{ex}}$. It is defined as the energy difference between the first excited state and the ground state, $\Delta_{\mathrm{ex}}=E_1-E_0$, in the many-body spectrum with fixed particle numbers and spins (for the 12A cluster, $N_{\uparrow}=N_{\downarrow}=6$). To illustrate its overall behavior, Fig.~\ref{fig_gap-sdw}(a) shows the ED results for $\Delta_{\mathrm{ex}}$ as functions of $U$ and $\Delta_{\mathrm{AB}}$ in a contour plot. Additionally, results at fixed $\Delta_{\mathrm{AB}}=4.0$ are displayed in Fig.~\ref{fig_gap4.0}(a). 

The contour plot of $\Delta_{\mathrm{ex}}$ [Fig.~\ref{fig_gap-sdw}(a)] reveals a region which, coinciding with the $C=1$ phase in the phase diagram [Fig.~\ref{fig_phase_ED}], is characterized by a very small excitation gap, with a maximum of only around $0.05$, spanning from the weak- to strong-coupling regimes of the phase diagram. At $\Delta_{\mathrm{AB}}=4.0$, the region lies within $U\in[7.04, 9.34]$ [Fig.~\ref{fig_gap4.0}(a)]. 

Figure~\ref{fig_gap-sdw}(b) shows the contour plot of the $z$-component AFM structure factor $S_{\mathrm{AFM}}^{zz}$ [Eq.~\eqref{eq_SDWzz_ED}]. Compared with the neighboring CDW and QSH ($C_s=1$) phases, the C1 phase exhibits stronger AFM correlations, consistent with the characteristics of the AFCI. Further details on the CDW and SDW correlations---particularly the AFM order along the $z$ axis versus that within the $xy$ plane---are presented in Fig.~\ref{fig_gap4.0}(a). All CDW and SDW structure factors display discontinuities at the phase boundaries of the C1 phase, signaling a phase transition. Importantly, within the C1 phase, AFM correlations along the $z$ axis dominate over those in the $xy$ plane, in contrast to the SDW phase where the later are favored, consistent with previous results~\cite{Rachel_2010,Rachel_2018,JiangK_2018}. We emphasize that the dominance of the AFM moment along the $z$ axis is a key feature associated with the emergence of the AFCI phase~\cite{JiangK_2018,Phillips_2025}. 

Additionally, in Fig.~\ref{fig_gap4.0}(a), we present the ED result on the single-particle charge gap $\Delta_{\mathrm{c}}$ alongside with the many-body excitation gap $\Delta_{\mathrm{ex}}$. The charge gap is defined as 
\begin{align}
\Delta_{\mathrm{c}}=&\frac{1}{2}\left[E_0\left(N_{\uparrow}+1,N_{\downarrow}+1\right)+E_0\left(N_{\uparrow}-1,N_{\downarrow}-1\right)\right] \notag \\ &-E_0(N),
\end{align}
where $E_0$ denotes the ground-state energy and $N=N_{\uparrow}+N_{\downarrow}$. We find that although $\Delta_{\mathrm{c}}$ remains finite in all four phases, it exhibits a pronounced dip at the CDW-C1 phase boundary. Moreover, within the C1 phase, $\Delta_{\mathrm{c}}$ increases with $U$ at a noticeably reduced rate. We note that in the DQMC simulations, the corresponding parameter regime is identified as an extended semimetallic region between the CDW and QSH phases based on compressibility calculation, signaling a tendency toward charge-gap closing~\cite{Phillips_2025}.  

To demonstrate that the C1 phase possesses topological characteristics distinguishing it from neighboring phases, we calculate the fidelity susceptibility~\cite{Zanardi_2006,Zanardi_2007a,Zanardi_2007b} with respect to TBC, which is defined as 
\begin{equation}
\chi_F=\frac{2}{N}\frac{1-\abs{\braket{\Psi_0(0)}{\Psi_0(\delta\theta)}}}{(\delta\theta)^{2}},
\label{eq_chiF}
\end{equation}
where $N$ is the lattice size, $\ket{\Psi_0(\delta\theta)}$ is the ground state under TBC with twist angle $\delta\theta$ ($\ket{\Psi_0(0)}$ corresponds to the ground state under PBC). The quantity $\chi_F$, according to the definition, measures the sensibility of the ground state to a small change in the boundary condition. In our calculation, we apply the spin-independent TBC~\eqref{eq_TBC_C} with $\delta\theta=0.01$ along the $\hat{\mathbf{x}}$ direction. The results for fixed $\Delta_{\mathrm{AB}}=4.0$ are shown in Fig.~\ref{fig_gap4.0}(b). We note that choosing $\delta\theta$ along the $\hat{\mathbf{y}}$ direction or applying the spin-dependent TBC of Eq.~\eqref{eq_TBC_Cs} yields qualitatively similar results. 

As shown in Fig.~\ref{fig_gap4.0}(b), peaks in the fidelity susceptibility $\chi_F$ appear at the boundaries between the CDW, C1, and QSH phases. This indicates that, being sandwiched between the topologically trivial CDW phase and the nontrivial QSH phase with $C_s=1$, the C1 phase must possess a distinct topological character. However, due to the TRS of the KMH Hamiltonian~\eqref{eq_ham}, which cannot be spontaneously broken in finite-size ED calculations, the Chern number $C$ obtained using spin-independent TBC~\eqref{eq_TBC_C} remains zero in all phases [Fig.~\ref{fig_cut4.0}(a)]. In contrast, with careful treatment, the spin Chern number $C_s$ [Eq.~\eqref{eq_cs}] extracted under spin-dependent TBC~\eqref{eq_TBC_Cs} takes the value $1/2$ in the C1 phase. In the next Sec.~\ref{subsec_Chern}, we present a modified algorithm that yields a nonzero Chern number $C=1$ for the C1 phase. 

Additionally, we notice the absence of a $\chi_F$ peak at the QSH-SDW transition in Fig.~\ref{fig_gap4.0}(b), where the SDW phase is topologically trivial. This absence arises from the lack of the gap closing under PBC, an issue we will return to in Sec.~\ref{subsec_gapclosure}.

The above analysis on the AFM order, topological invariants, and the fidelity susceptibility with respect to TBC indicates that the C1 phase exhibits the essential characteristics of the AFCI. The excitation gap in this phase is found to be very small. While the charge gap stays finite, it reaches a minimum at the CDW-C1 phase boundary and grows only weakly with increasing $U$, reminiscent of the semimetallic scenario proposed in Ref.~\cite{Phillips_2025}. Overall, the question of the existence and stability of the AFCI in the KMH model remains open and requires further evidence and cross-validation from different methods.

\subsection{Modified algorithm for the Chern number} \label{subsec_Chern}

\begin{figure}
\centering
\includegraphics[width=0.47\textwidth]{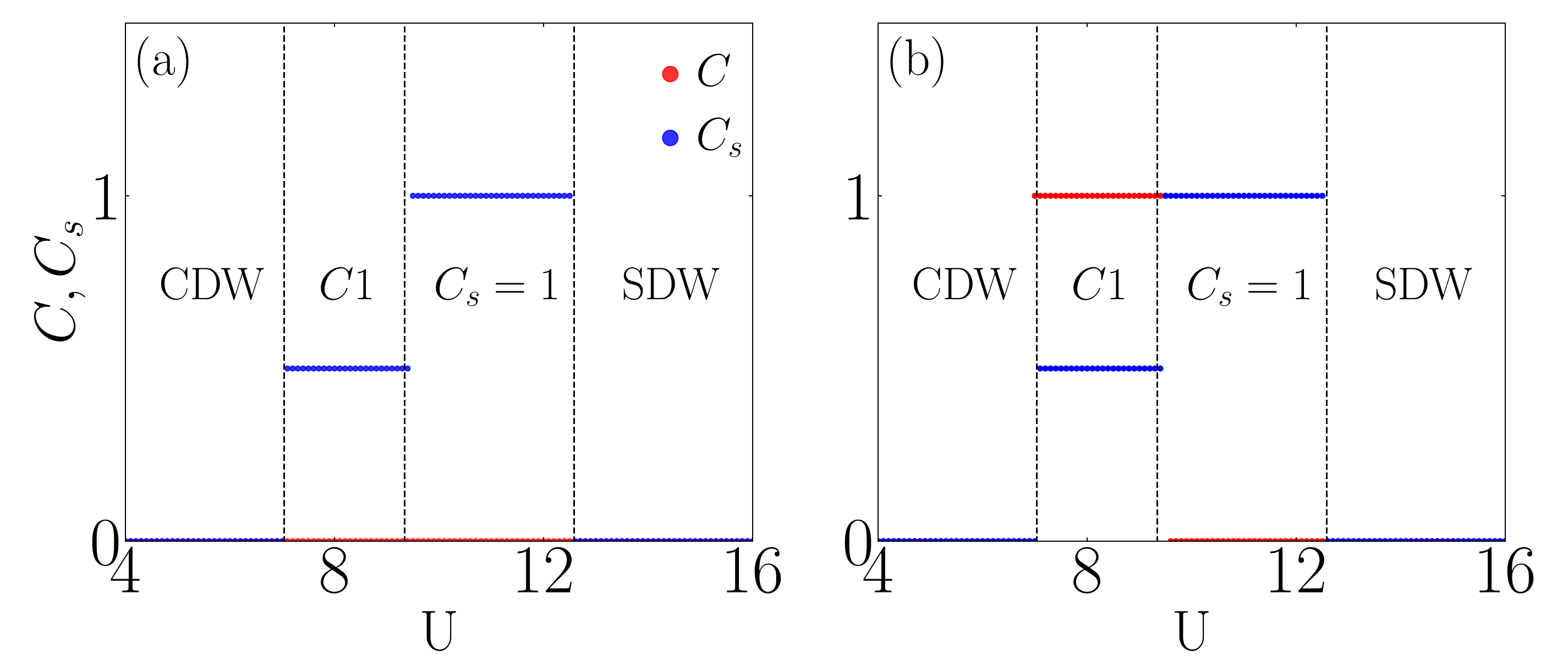}
\caption{Evaluation of the Chern number $C$ and spin Chern number $C_s$ as functions of $U$ at fixed $\Delta_{\mathrm{AB}} = 4.0$. Panel (a) shows the results obtained using conventional methods, while panel (b) presents those obtained with the modified algorithm described in the text.}
\label{fig_cut4.0}
\end{figure}

In Fig.~\ref{fig_cut4.0}(a), we use the spin-independent and spin-dependent TBCs, defined in Eqs.~\eqref{eq_TBC_C} and \eqref{eq_TBC_Cs}, to compute the Chern number and the spin Chern number on the discretized $\bm{\theta}$ space~\cite{Fukui_2005}, at fixed $\Delta_{\mathrm{AB}}$. The spin Chern number $C_s$ [Eq.~\eqref{eq_cs}] is found to be $1/2$ in the C1 phase. This result naturally suggests that the C1 phase corresponds to the AFCI, with one spin sector being topologically nontrivial while other remains trivial. In this case, TRS should be broken, and the resulting Chern number in the C1 phase should equal $1$. However, as shown in Fig.~\ref{fig_cut4.0}(a), this does not occur in our ED calculation. This outcome sounds reasonable since the spontaneous TRS breaking cannot occur in finite-size systems governed by a TRS Hamiltonian, and all eigenstates necessarily respect TRS. Consequently, one might expect that the Chern number remains zero in all phases~\cite{Moore_2007}. 

However, for the C1 phase, recall that the excitation gap obtained under PBC is found to be ubiquitous small. We therefore surmise that, with TBC applied---which breaks translational invariance as well as TRS {\em at the boundary}---the adiabaticity condition (i.e., the requirement of a finite gap) may fail, leading to a violation of the admissibility condition for applying Eq.~\eqref{eq_C} and thereby invalidating the standard numerical approach.

\begin{figure}
\centering
\includegraphics[width=0.45\textwidth]{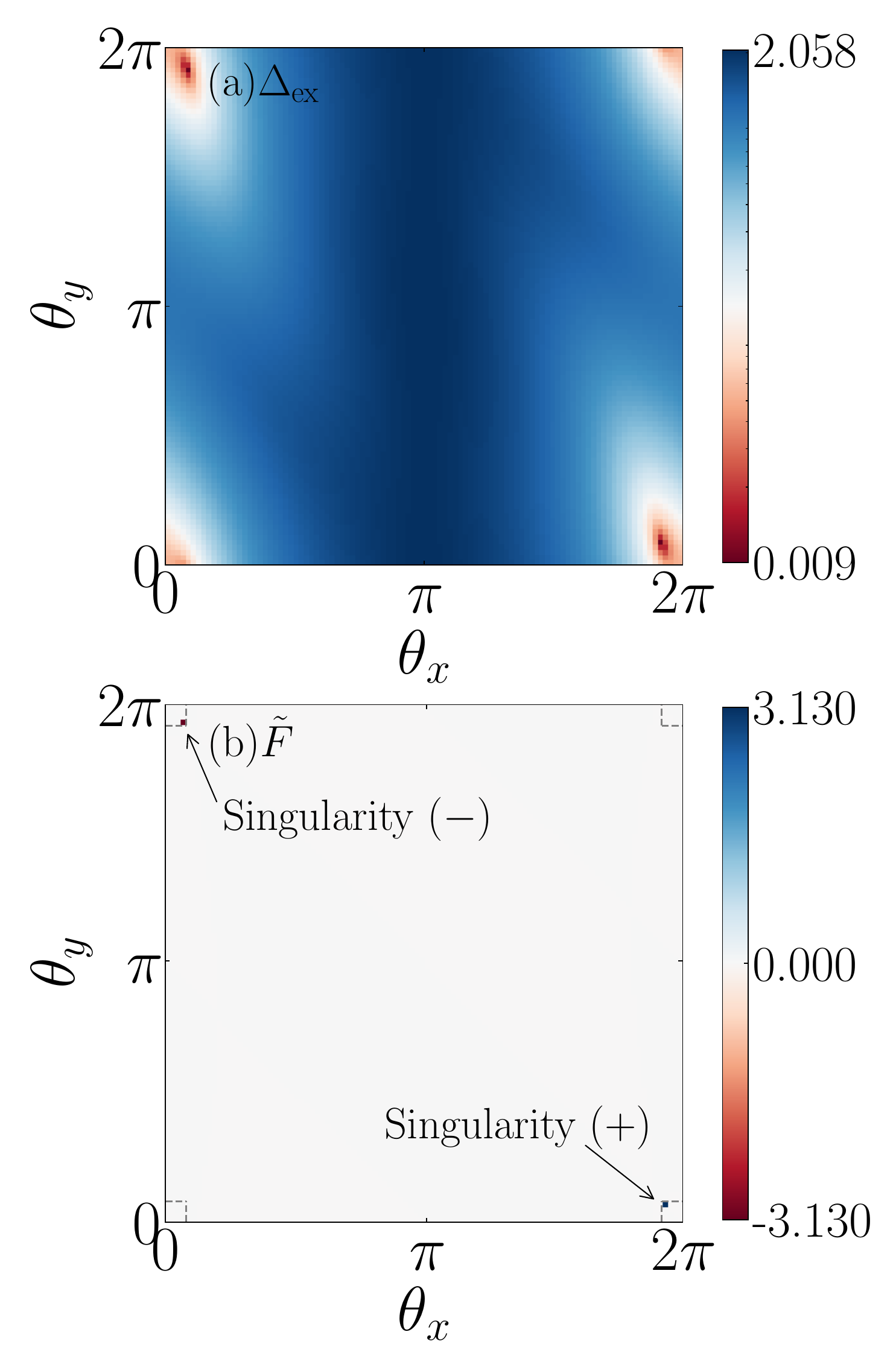}
\caption{(a) Excitation gap $\Delta_{\mathrm{ex}}$ and (b) lattice field strength $\tilde{F}$ [Eq.~\eqref{eq_Ftilde}] as functions of twist angles $(\theta_x, \theta_y)$ under spin-independent TBCs. Dashed boundaries in (b) denote the swap region, where the Berry curvature is evaluated using the first exited state of $\hat{H}(\bm{\theta})$ instead of the ground state. See the main text for details. Parameters: $\Delta_{\mathrm{AB}}=4.0$ and $U=8.0$.}
\label{fig_U8}
\end{figure}

To examine the possible breakdown of adiabaticity, we calculate the excitation gap $\Delta_{\mathrm{ex}}$ and the lattice field strength $\tilde{F}$ on a fine $100\times 100$ grid over the toroidal $\bm{\theta}$ space. The results for a representative parameter set, $\left(\Delta_{\mathrm{AB}},\,U\right)=(4.0,\,8.0)$, are shown in Fig.~\ref{fig_U8}. Under this discretization, $\tilde{F}$ appears to develop a pair of singularity-like values (with magnitude close to $\pi$ but {\em opposite in sign}) at $\left(\theta_x, \theta_y\right) = 2\pi\left(\frac{4}{100}, \frac{96}{100}\right)$ and $\left(\theta_x, \theta_y\right) = 2\pi\left(\frac{96}{100}, \frac{4}{100}\right)$. The shorter distance between the two points, measured along the antidiagonal, is $0.08\sqrt{2}\times 2\pi$. Away from them, $\tilde{F}$ tends towards zero. Correspondingly, the excitation gap $\Delta_{\mathrm{ex}}$ shows minima of order $\sim 0.01$ at the same locations. These points appear embedded in a continuous, isolated region of $\bm{\theta}$ space where $\Delta_{\mathrm{ex}}$ remains small [visible as the red-shifted area in Fig.~\ref{fig_U8}(a)], covering roughly a little more than $8\times 8$ plaquettes in total.

\begin{figure}
\centering
\includegraphics[width=0.47\textwidth]{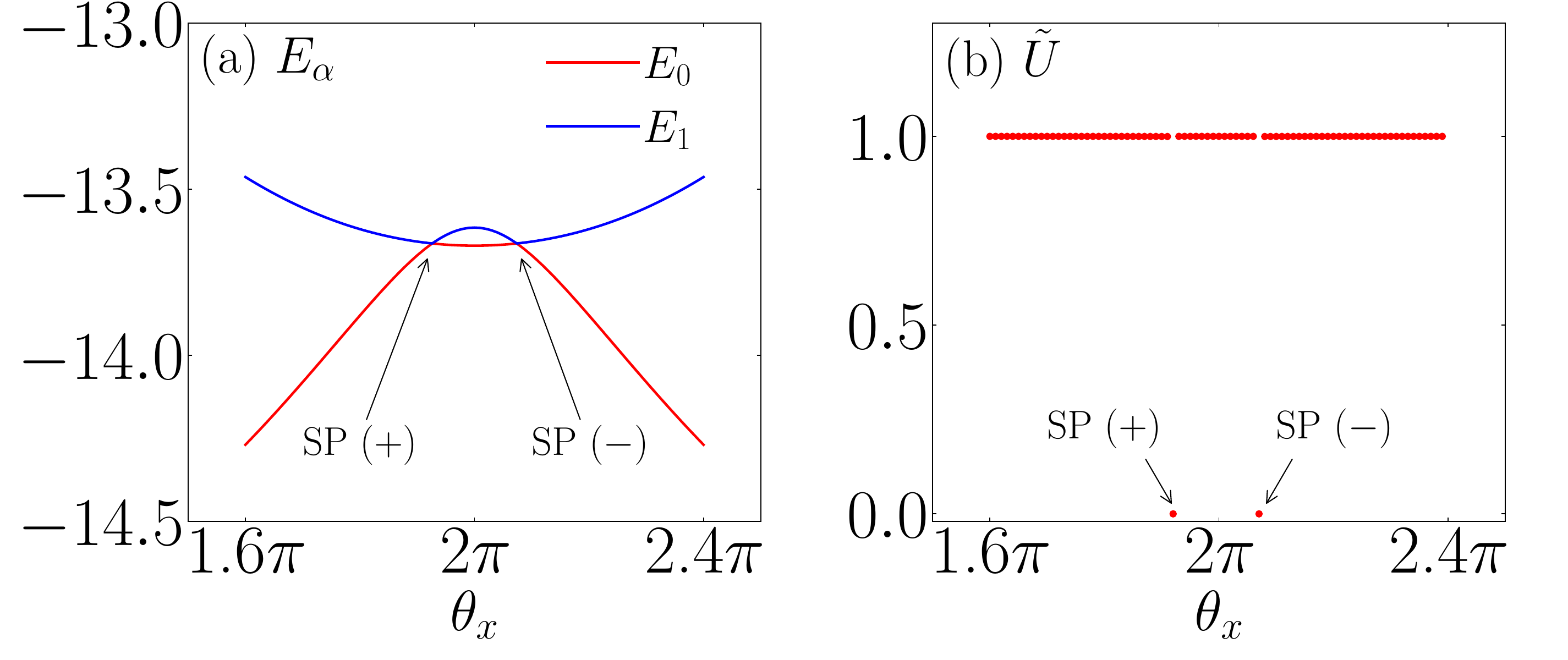}
\caption{(a) Ground-state energy $E_0$ and first excitation energy $E_1$, and (b) magnitude of the unnormalized $\mathrm{U}(1)$ link $\tilde{U}$, defined as $\tilde{U}\left(\theta_{\ell},\delta\theta_{\ell}\right)=\abs{\braket{\Psi\left(\theta_{\ell}\right)}{\Psi\left(\theta_{\ell}+\delta\theta_{\ell}\right)}}$, plotted as functions of $\theta_x$ along the antidiagonal line in $\bm{\theta}$ space. ``SP'' denotes a singular point. Parameters: $\Delta_{\mathrm{AB}}=4.0$ and $U=8.0$.}
\label{fig_thetaY}
\end{figure}

The above observations suggest that at the identified twist angles associated with the singularities in $\tilde{F}$, {\em true} level crossings, potentially responsible for the breakdown of adiabaticity, may occur~\cite{Varney_2011}. To confirm this, we compute the ground-state energy $E_0$ and the first excitation energy $E_1$ along the antidiagonal direction as functions of $\theta_x$. As shown in Fig.~\ref{fig_thetaY}(a), level crossings---rather than avoided ones---occur at the two singular points, and between them the ground state and the first excited state exchange their positions. This phenomenon is further corroborated by the behavior of the magnitude of the unnormalized $\mathrm{U}(1)$ link variable $\tilde{U}$ along the antidiagonal, shown in Fig.~\ref{fig_thetaY}(b). At the two crossing points, $\tilde{U}$ exhibits sharp dips toward zero, signaling the level crossings, while elsewhere it remains close to unity. 

Such breakdown of adiabaticity, arising from level crossings between the ground and the first excited states and signaled by the appearance of paired singularities in the lattice field strength $\tilde{F}$, are ubiquitously observed in the C1 phase. These paired singularities consistently lie along the antidiagonal direction and are symmetrically positioned about the origin in $\bm{\theta}$ space. Their separation varies across the phase diagram, reaching up to about 8\% of the total antidiagonal length, while becoming smaller in the vicinity of the C1 phase boundary. 



With this picture in mind, and noting that the level crossings are generally confined to a narrow region in $\bm{\theta}$ space, we propose a simple---albeit crude---remedy. Specifically, a swap region encompassing the singularities [for instance, in Fig.~\ref{fig_U8}(b), an $8\times 8$ plaquette region enclosed by dashed boundaries and centered at the origin]. Within this region, when evaluating the $\mathrm{U}(1)$ link [Eq.~\eqref{eq_link}], we use the first excited state of of $\hat{H}(\bm{\theta})$ in place of the ground state. This substitution allows us to more smoothly follow the evolution of the targeted wavefunction across the $\bm{\theta}$ space, thereby mitigating the severity of singularities in the curvature estimation.

Figure~\ref{fig_cut4.0}(b) shows the result of the modified Chern number calculation at $\Delta_{\mathrm{AB}}=4.0$, which yields $C=1$ in the C1 phase. The numerical procedure remains stable within this phase, provided that the square region where the first excited state is used encompasses the singularities; this region may exceed the minimal size and extend up to $15\times 15$ plaquettes on the $100\times 100$ grid in $\bm{\theta}$ space. In the other phases, such level crossings are absent, and the standard algorithm suffices.

\subsection{Phase diagram of the KMH model with mass imbalance}
\label{subsec_massimbalance}

In the previous two subsections, we presented evidence for the existence of the AFCI phase in the KMH model based on AFM orders, fidelity susceptibility, and (spin) Chern number calculations. In particular, we propose a modified algorithm for Chern number evaluation that remains valid even when the gap-opening condition under TBCs is not satisfied, successfully yielding $C=1$ for the C1 phase. 

In this subsection, we examine the phase diagram of the KMH Hamiltonian with a finite mass imbalance in the NN hoppings, defined as
\begin{equation}
\hat{H}_{t_1}=-\sum\limits_{\vev{ij},\sigma}\left(t_{1}+\sigma_{z} \delta\right){c}_{i\sigma}^{\dagger}{c}_{j\sigma},
\label{eq_imbalance}
\end{equation}  
where $\delta$ denotes the imbalance strength.

The motivation of introducing mass imbalance~\cite{Dao_2012,Vanhala_2016,DuL_2017,DuL_2023} is to probe the system's response to explicit TRS breaking in a controlled fashion. It can be expected that the spin-dependent modulation on NN hoppings can enhance the susceptibility to the Hubbard-induced AFM perturbations, thereby facilitating the emergence and stabilization of the AFCI phase.

\begin{figure}
\centering
\includegraphics[width=0.45\textwidth]{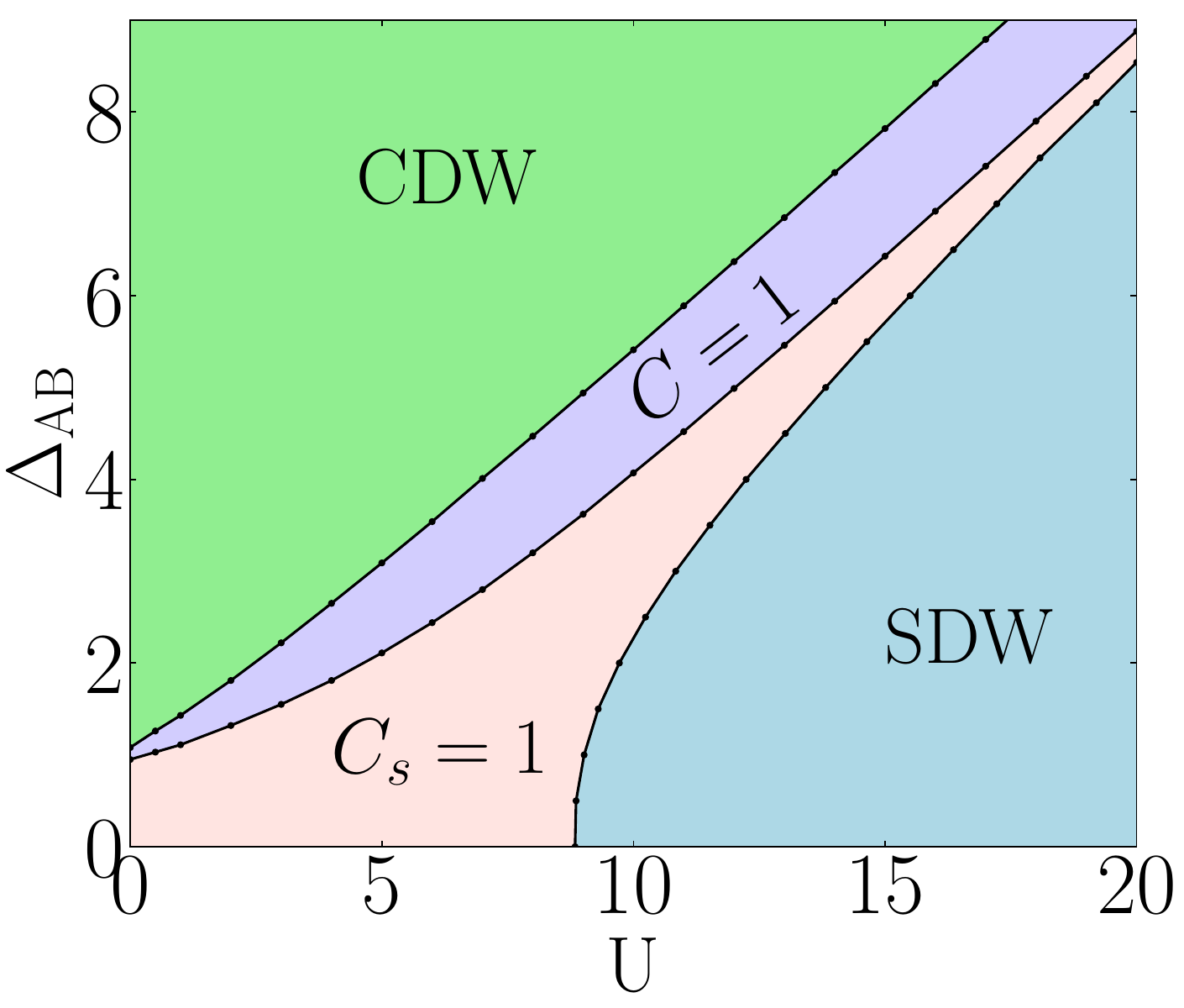}
\caption{ED phase diagram of the KMH Hamiltonian~\eqref{eq_ham} with mass imbalance $\delta=0.1$~\eqref{eq_imbalance}. Notably, without interaction ($U = 0$), the $C=1$ phase appears in the range $\Delta_{\mathrm{AB}}\in[0.95,\,1.08]$.}
\label{fig_imbalance}
\end{figure}

Figure~\ref{fig_imbalance} shows the ED phase diagram of the KMH model with a finite mass imbalance $\delta=0.1$. The overall structure closely resembles that of the balanced case (Fig.~\ref{fig_phase_ED}), but the $C=1$ region expands, while the QSH phase is slightly reduced. Remarkably, even in the noninteracting limit ($U=0$), the AFCI phase already appears within $\Delta_{\mathrm{AB}}\in[0.95,\,1.08]$. This follows from the spin dependence of the resulting effective staggered potential induced by the mass imbalance.

With mass imbalance present, the Chern number of the AFCI phase can be computed using the standard procedure since no level crossings occur under TBCs and adiabatic continuity is preserved throughout $\bm{\theta}$ space. The robustness of the $C=1$ phase under finite mass imbalance suggests that introducing such an imbalance offers a practical route to experimentally realize and stabilize the AFCI phase in KMH-related systems.

\subsection{Gap closing at the QSH-SDW phase boundary}
\label{subsec_gapclosure}

From the ED results on the 12A cluster presented in Sec.~\ref{subsec_C1}, we observed that the excitation gap remains open under PBC at the QSH-SDW phase boundary [Figs.~\ref{fig_gap-sdw}(a)]. Consequently, the structures factors vary smoothly [Fig.~\ref{fig_gap4.0}(a)], and the fidelity susceptibility $\chi_F$ exhibits no peak at the transition point [Fig.~\ref{fig_gap4.0}(b)]. However, the transition from the QSH phase with spin Chern number $C_s=1$ to the topologically trivial SDW phase is, by nature, a topological transition. Thus, with the change of the topological invariant, the many-body gap (or the band gap in the noninteracting case) must close at the transition point, provided that the system's symmetry is preserved throughout the process~\cite{Nagaosa_2013}. 

\begin{figure}
\centering
\includegraphics[width=0.5\textwidth]{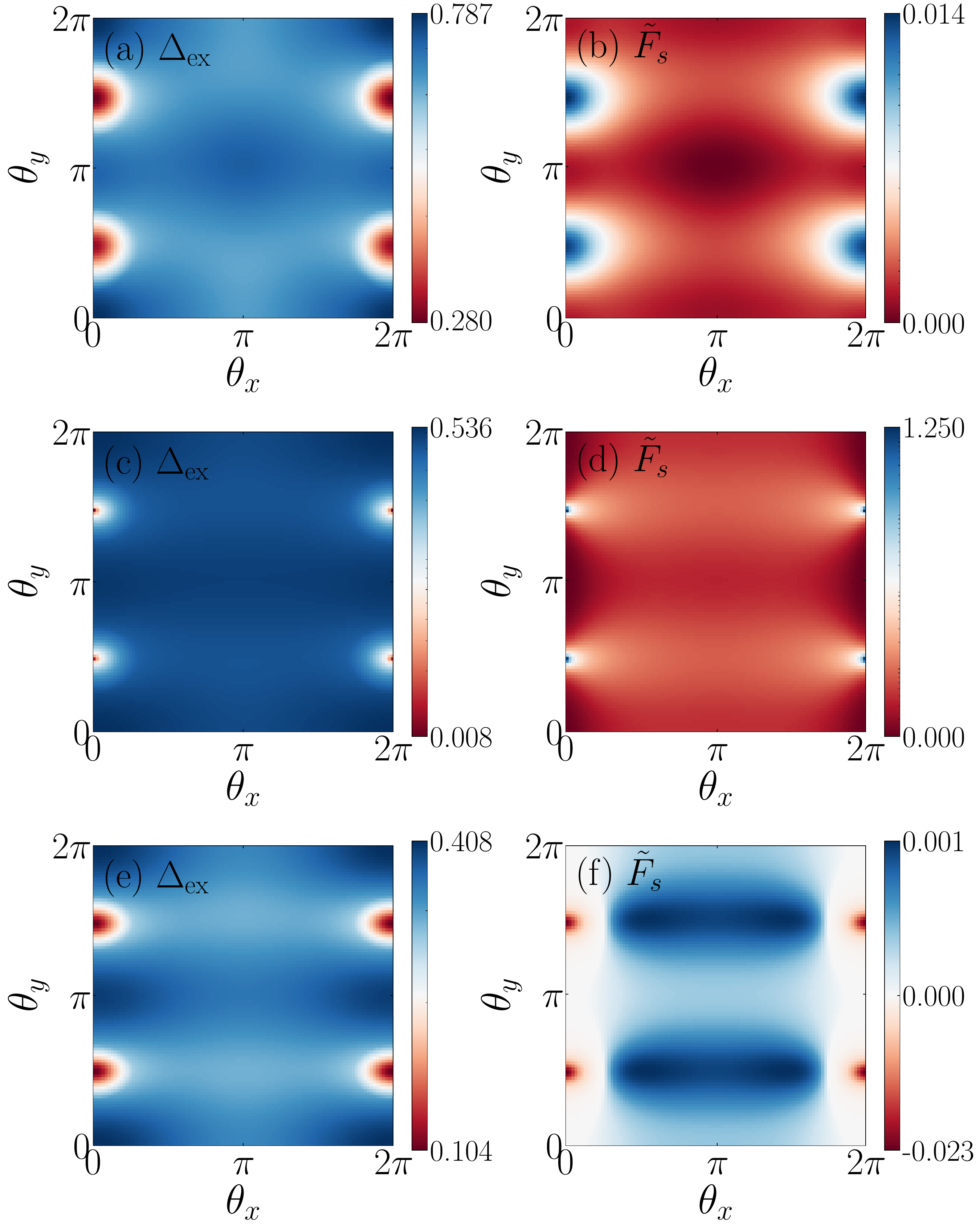}
\caption{Excitation gap $\Delta_{\text{ex}}$ and lattice field strength $\tilde{F}_s$ as functions of the twisted angles $\left(\theta_x,\,\theta_y\right)$ under spin-dependent TBCs~\eqref{eq_TBC_Cs}, with sublattice potential fixed at $\Delta_{\mathrm{AB}} = 0.5$. (a,b) Results at $U = 7.0$ (QSH phase); (c,d) $U = 8.95$ (near the phase boundary); (e,f) $U = 11.0$ (SDW phase). Calculations are performed on a $100\times 100$ grid in $\bm{\theta}$ space.}
\label{fig_cut0.5}
\end{figure}

To clarify the issue, we examine the excitation gap $\Delta_{\mathrm{ex}}$ and the lattice field strength $\tilde{F}_s$ under spin-dependent TBCs~\eqref{eq_TBC_Cs} along a horizontal cut of the ED phase diagram [Fig.~\ref{fig_phase_ED}] at fixed $\Delta_{\mathrm{AB}}=0.5$. As the interaction strength $U$ increases, the system undergoes a topological phase transition from the QSH to the SDW phase. Representative results for $U=7.0$ (QSH phase), $U=8.95$ (near the transition), and $U=11.0$ (SDW phase) are shown in Fig.~\ref{fig_cut0.5}.

From Fig.~\ref{fig_cut0.5}, we see that within both the QSH and SDW phases the system remains well gapped across the entire $\bm{\theta}$ space [Figs.~\ref{fig_cut0.5}(a) and \ref{fig_cut0.5}(e)], and the lattice field strength $\tilde{F}_s$ remains smooth, small in magnitude, and free of singularities. In contrast, at $U=8.95$, close to the critical point, two local minima of the excitation gap $\Delta_{\mathrm{AB}}$ emerge at $\left(\theta_x,\,\theta_y\right)=2\pi\left(0,\,\frac{24}{100}\right)$ and $2\pi\left(0,\,\frac{74}{100}\right)$ [Fig.~\ref{fig_cut0.5}(c)]. Correspondingly, $\tilde{F}_s$ develops singularity-like features with magnitude of order unity at these positions [Fig.~\ref{fig_cut0.5}(d)]. 

\begin{figure}
\centering
\includegraphics[width=0.4\textwidth]{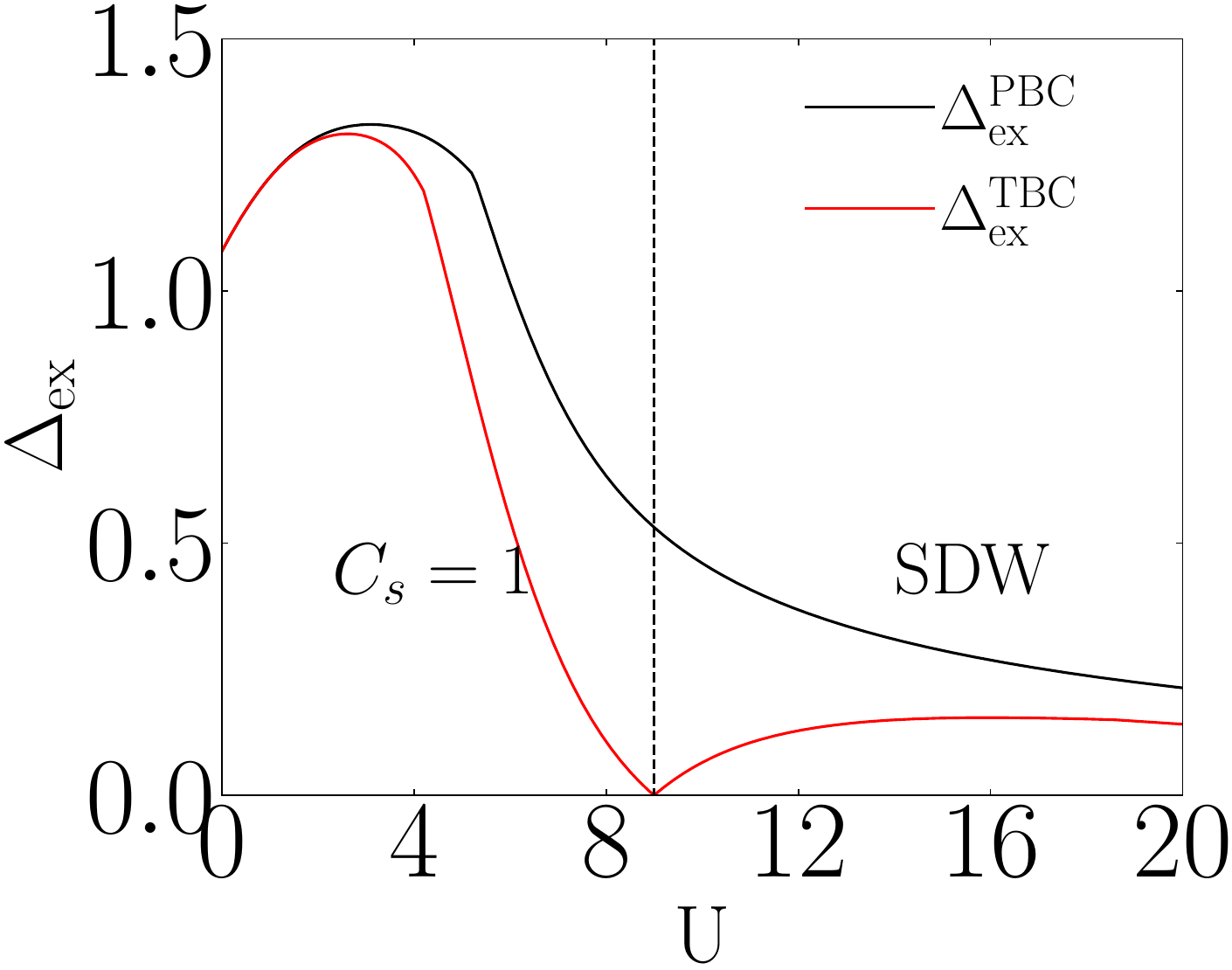}
\caption{Evolution of the excitation gaps as functions of $U$ under PBC and spin-dependent TBCs~\eqref{eq_TBC_Cs}, denoted by $\Delta_{\mathrm{ex}}^{\mathrm{PBC}}$ and $\Delta_{\mathrm{ex}}^{\mathrm{TBC}}$, respectively. The values of $\Delta_{\mathrm{ex}}^{\mathrm{TBC}}$ are obtained as the minimum excitation gap over the entire $\bm{\theta}$ space. The sublattice potential is fixed at $\Delta_{\mathrm{AB}}=0.5$.}
\label{fig_gap0.5}
\end{figure}

Figure~\ref{fig_gap0.5} presents the evolution of the excitation gaps as functions of $U$ under PBC and spin-dependent TBCs, respectively. The quantity $\Delta_{\mathrm{ex}}^{\mathrm{TBC}}$, defined as the minimum excitation gap over the full $\bm{\theta}$ space, is observed to approach zero at the transition point. We further find that the deviation between $\Delta_{\mathrm{ex}}^{\mathrm{PBC}}$ and $\Delta_{\mathrm{ex}}^{\mathrm{TBC}}$ is maximal at the QSH-SDW phase boundary and is suppressed in both weak- and strong-coupling limits, which may indicate enhanced quantum fluctuations and pronounced finite-size effects near the transition.

These observations suggest that, to accommodate the change of topological numbers across the phase transition, the many-body excitation gap is expected to close at certain points in $\bm{\theta}$ space at the transition. In the present case, the closing points are distinct from the zeros corresponding to PBC. The apparent absence of gap closing under PBC leads to smooth behavior in the structure factors [Fig.~\ref{fig_gap4.0}(a)] and the absence of any pronounced peak in the fidelity susceptibility [Fig.~\ref{fig_gap4.0}(b)]. A similar phenomenon was reported in ED calculation of the HH model with sublattice potentials on the same cluster~\cite{Shao_2023}, where at the phase boundary between the AFCI and $C=2$ phases the gap-closing point was close to, but distinct from, the origin (PBC). In contrast, for the KMH model studied here, the closing occurs at shifted points in $\bm{\theta}$ space. We note that the DQMC study of the compressibility indicates that the {\em charge gap} remains finite during the QSH-SDW transition~\cite{Phillips_2024,Phillips_2025}. 



\section{Mean-field results}
\label{sec_results_MF}

In this section, we briefly summarize the Hartree-Fock MF results for the KMH model~\eqref{eq_ham}. Specific attention is paid to the evaluation of the (spin) Chern number in terms of the effective Dirac masses in the MF Hamiltonian~\cite{HeWX_2024,Wang_2024}, which provides an intuitive understanding of the emergence of the AFCI phase at the MF level.

The MF Hamiltonian of the KMH model with sublattice potential $\Delta_{\mathrm{AB}}$ in the momentum space takes the form
\begin{equation}
\hat{H}_{\mathrm{MF}}=\hat{H}_{0}+\sum_{\mathbf{k}}\hat{\psi}_{\mathbf{k}}^{\dag}
\begin{pmatrix} \varepsilon_{\uparrow}^{\mathrm{A}} & 0 & \varepsilon_{\uparrow\downarrow}^{\mathrm{A}} & 0 \\ 0 & \varepsilon_{\uparrow}^{\mathrm{B}} & 0 & \varepsilon_{\uparrow\downarrow}^{\mathrm{B}} \\ \varepsilon_{\uparrow\downarrow}^{\ast \mathrm{A}} & 0 & \varepsilon_{\downarrow}^{\mathrm{A}} & 0   \\ 0 & \varepsilon_{\uparrow\downarrow}^{\ast \mathrm{B}} &  0 & \varepsilon_{\downarrow}^{\mathrm{B}} \\
\end{pmatrix}\hat{\psi}_{\mathbf{k}},
\label{eq_ham_MF}
\end{equation}
where $\hat{H}_{0}$ is the noninteracting Hamiltonian~\eqref{eq_ham_0} and $\hat{\psi}^{\dag}_{\mathbf{k}}=\left[a^{\dag}_{\mathbf{k}\uparrow}\ b^{\dag}_{\mathbf{k}\uparrow}\ a^{\dag}_{\mathbf{k}\downarrow}\ b^{\dag}_{\mathbf{k}\downarrow}\right]$. Here, $a$ ($a^{\dagger}$) and $b$ ($b^{\dagger}$) denote the annihilation (creation) operators for electrons on the $A$ and $B$ sublattices, respectively.

The matrix elements in the second term of $\hat{H}_{\mathrm{MF}}$~\eqref{eq_ham_MF} arise from the MF decoupling of the interaction term $H_{\mathrm{I}}$~\eqref{eq_ham_I}, and are given by
\begin{align}
&\varepsilon_{\sigma}^{\mathrm{A}/\mathrm{B}}=U n^{\mathrm{A}/\mathrm{B}}_{\bar{\sigma}},\notag \\
&\varepsilon_{\uparrow\downarrow}^{\mathrm{A}}=-\frac{U}{N}\sum_{\mathbf{q}}\vev{a^{\dag}_{\mathbf{q}\downarrow}a_{\mathbf{q}\uparrow}}_{\mathrm{MF}},\notag  \\
&\varepsilon_{\uparrow\downarrow}^{\mathrm{B}}=-\frac{U}{N}\sum_{\mathbf{q}}\vev{b^{\dag}_{\mathbf{q}\downarrow}b_{\mathbf{q}\uparrow}}_{\mathrm{MF}},
\end{align}
where $n^{\mathrm{A}}_{\sigma}=\frac{1}{N}\sum_{\mathbf{q}}\vev{a^{\dag}_{\mathbf{q}\sigma}a_{\mathbf{q}\sigma}}_{\mathrm{MF}}$ (and similarly for $n^{\mathrm{B}}_{\sigma}$), and $\bar{\sigma}$ denotes the spin opposite to $\sigma$. The averages $\vev{\cdots}$ are evaluated in the grand-canonical ensemble with respect to the MF Hamiltonian at sufficiently low temperature, with the chemical potential determined self-consistently at fixed filling.

Convergence of the MF solution is typically monitored through the free energy. In the present calculation, the MF parameters converge already on a $30\times 30$ lattice, exhibiting only weak size dependence. Further details can be found, e.g., in Refs.~\cite{Shao_2021,HeWX_2024,Wang_2024}.

\begin{figure}
\centering
\includegraphics[width=0.45\textwidth]{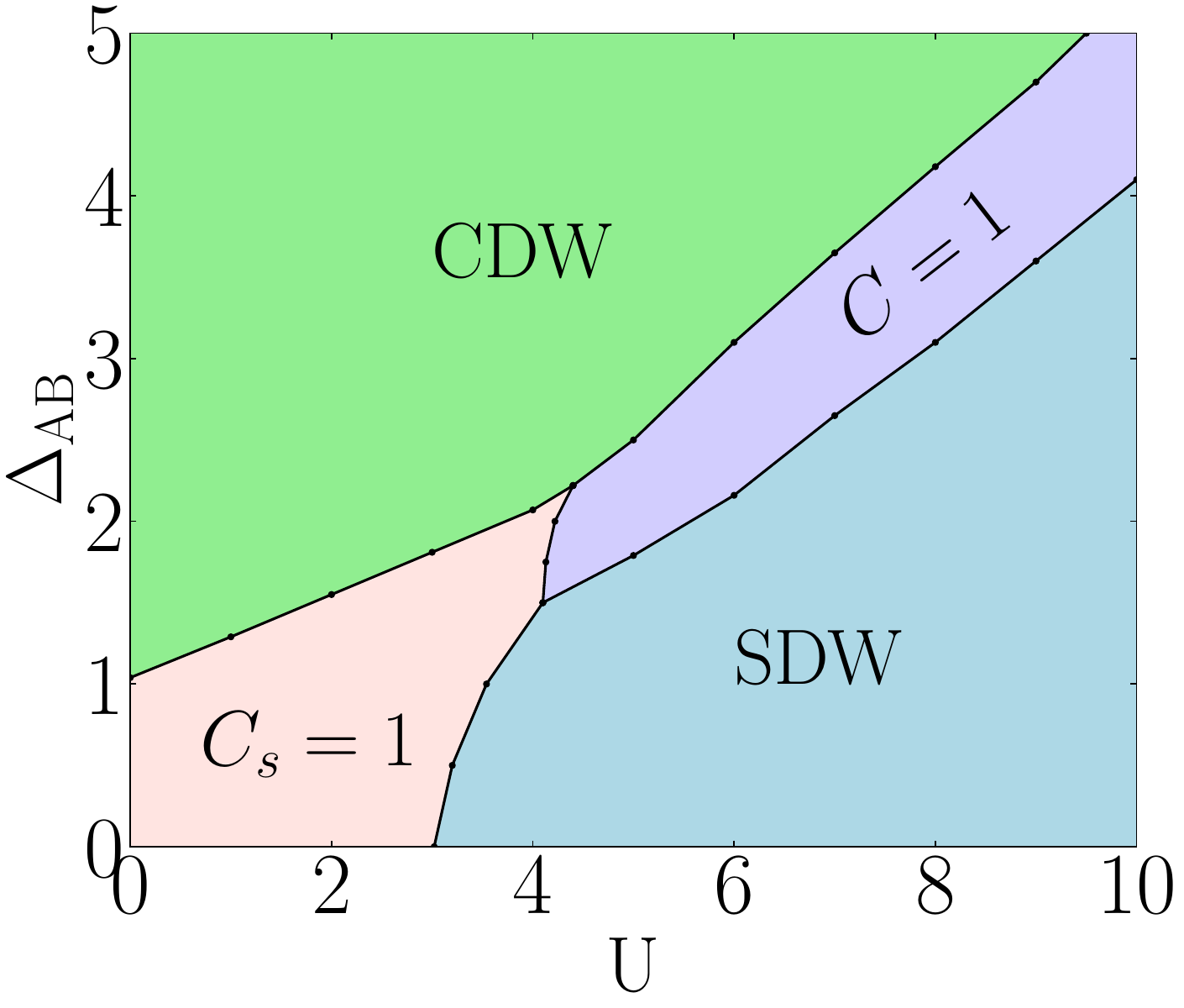}
\caption{MF phase diagram of the KMH model~\eqref{eq_ham} in the $(U,\Delta_{\mathrm{AB}})$ plane with fixed $t_2=0.2$, obtained on a $30\times 30$ lattice.}
\label{fig_phase_MF}
\end{figure}

Figure~\ref{fig_phase_MF} shows the MF phase diagram of the KMH model~\eqref{eq_ham} at fixed $t_2=0.2$, obtained from the MF order parameters (SDW and CDW; not shown), together with the topological invariants ($C$ and $C_s$) of the MF Hamiltonian~\eqref{eq_ham_MF}. The procedure follows standard practice, and the results are consistent with previous studies~\cite{JiangK_2018}. In the following, we focus on the emergence of the AFCI phase in the KMH model at the MF level.

It is well known that, analogous to the Haldane model~\cite{Haldane_1988}, the band topology of the KM model with only intrinsic SOC [Eq.~\eqref{eq_ham_0}] is determined by the signs of the Dirac masses at the valleys $K$ and $K'$. These take the form $\pm 3\sqrt{3}t_2\sigma_z$, where the sign distinguishes the two valleys and $\sigma_z$ the two spin components. The opposite signs of the Dirac masses at $K$ and $K'$ yield a nonzero Chern number for the occupied band. The presence of TRS in the KM model enforces opposite Chern numbers for the two spin sectors, e.g., $C_{\uparrow}=1$ and $C_{\downarrow}=-1$, giving a vanishing total Chern number but a finite spin Chern number $C_s=1$ [Eq.~\eqref{eq_cs}]. 

Introducing a sublattice potential modifies the mass term to $m_{\sigma}=\Delta_{\mathrm{AB}}\pm 3\sqrt{3}t_2\sigma_z$. With sufficiently large sublattice potential (to be precise, when $\Delta_{\mathrm{AB}}>3\sqrt{3}t_2$), the sign difference of the Dirac masses between the two valleys disappears for each spin component, eliminating band inversion and driving the system into a topologically trivial band-insulating phase. 

The Hubbard interaction [Eq.~\eqref{eq_ham_I}] further renormalizes the mass term at the MF level. Examining the second term of the MF Hamiltonian~\eqref{eq_ham_MF}, one sees that the Hartree-Fock decoupling produces momentum-independent matrix elements. Among these, the diagonal Hartree terms yield an effective spin-dependent sublattice potential, given by 
\begin{equation}
\delta_{\mathrm{MF}}^{\sigma}
= \Delta_{\mathrm{AB}} + \frac{1}{2}\left(\varepsilon_{\sigma}^{\mathrm{A}}-\varepsilon_{\sigma}^{\mathrm{B}}\right)=\Delta_{\mathrm{AB}}+ \frac{U}{2}\left(n^{\mathrm{A}}_{\bar{\sigma}} - n^{\mathrm{B}}_{\bar{\sigma}}\right).
\label{eq_delta_sigma}
\end{equation}
Accordingly, the effective Dirac mass becomes
\begin{equation}
m_{\sigma}=\delta_{\mathrm{MF}}^{\sigma}\pm 3\sqrt{3}t_2\sigma_z.
\label{eq_msigma}
\end{equation}
Note that Eq.~\eqref{eq_delta_sigma} implies that the difference between the effective sublattice potentials of the two spin components is given by
\begin{equation}
\Delta\delta_{\mathrm{MF}}^{\uparrow\downarrow}=\frac{U}{2}\left(\sigma^{\mathrm{B}}_z-\sigma^{\mathrm{A}}_z\right), 
\label{eq_delta_delta}
\end{equation}
where $\sigma^{\mathrm{A}/\mathrm{B}}_z=n^{\mathrm{A}/\mathrm{B}}_{\uparrow}-n^{\mathrm{A}/\mathrm{B}}_{\downarrow}$.

This observation indicates that in the presence of AFM order along the $z$ direction, which breaks spin-rotation symmetry, an additional phase may arise between the QSH and trivial phases. Specifically, if $\delta_{\mathrm{MF}}^{\sigma}$ is large enough to trivialize one spin component but not the other, a phase with net Chern number $C=1$ emerges. This phase, characterized by correlation-driven AFM order along $z$, is known to be the AFCI phase~\cite{He_2011,Vanhala_2016,JiangK_2018}.

\begin{figure}
\centering
\includegraphics[width=0.45\textwidth]{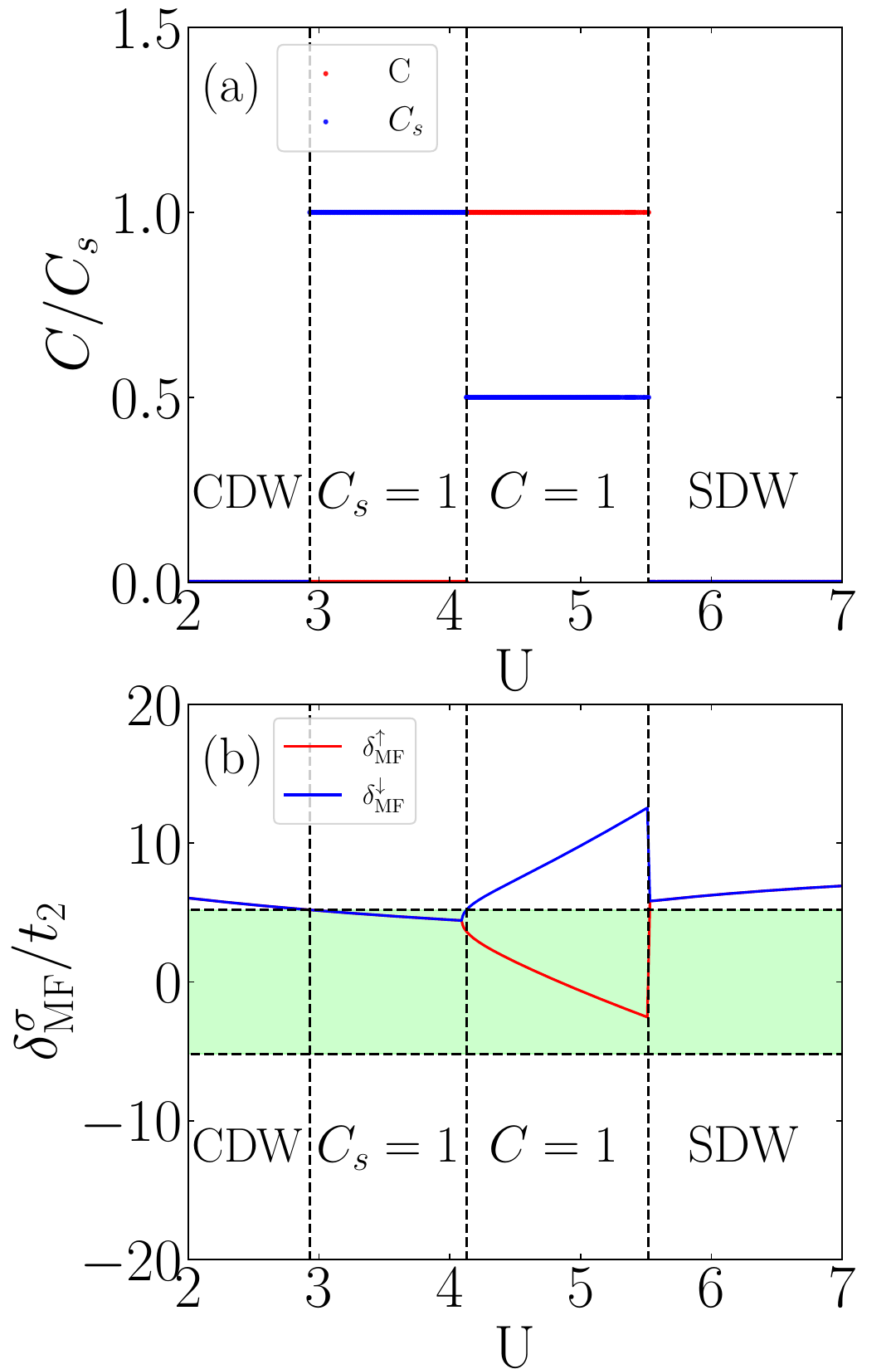}
\caption{MF analysis of the topological phase transitions in the KMH model~\eqref{eq_ham} for fixed $\Delta_{\mathrm{AB}} = 1.8$. As the Hubbard interaction $U$ increases, the system undergoes successive transitions CDW $\to$ QSH ($C_s=1$) $\to$ AFCI ($C=1$) $\to$ SDW, with critical values $U=2.93$, $4.13$ and $5.52$, respectively. Vertical dashed lines in both panels mark the phase boundaries obtained from the topological invariants $C$ and $C_s$, evaluated via conventional methods in terms of the MF Hamiltonian~\eqref{eq_ham_MF}. In (a), the solid lines for the (spin) Chern number $C$ ($C_s$) are obtained from the effective Dirac mass~\eqref{eq_msigma}, as explained in the text. Panel (b) shows the effective spin‐resolved sublattice potential $\delta_{\mathrm{MF}}^{\sigma}$~\eqref{eq_delta_sigma} as a function of $U$ for both spin species. The shaded region indicates the parameter range where $|\delta_{\mathrm{MF}}^{\sigma}/t_2| < 3\sqrt{3}$.}
\label{fig_mass}
\end{figure}

The correlation-driven topological phase transitions at the MF level are illustrated in Fig.~\ref{fig_mass}. For fixed $\Delta_{\mathrm{AB}}=1.8$, the figure shows the MF results for the spin-resolved effective sublattice potential as functions of $U$, together with the corresponding topological invariants $C$ and $C_s$ obtained from the effective Dirac mass [Eq.~\eqref{eq_msigma}]. 

As we observe that at fixed $\Delta_{\mathrm{AB}}=1.8$, increasing $U$ drives a sequence of topological phase transitions CDW $\to$ QSH $\to$ AFCI $\to$ SDW. In the topologically trivial phases (CDW and SDW), the effective sublattice potentials satisfy $\abs{\delta_{\mathrm{MF}}^{\uparrow/\downarrow}/t_2}>3\sqrt{3}$ [outside the shaded region in Fig.~\ref{fig_mass}(b)], indicating that the MF occupied bands of both spin components are trivial. In the QSH phase, the condition reverses to $\abs{\delta_{\mathrm{MF}}^{\uparrow/\downarrow}/t_2}<3\sqrt{3}$ [inside the shaded region as shown in Fig.~\ref{fig_mass}(b)], so that both occupied bands are topologically nontrivial. Because the sign of the $t_2$ term flips between spins in the effective Dirac mass~\eqref{eq_msigma}, the two spin bands carry opposite Chern numbers, giving $C_s=1$. 

In contrast, in the $C=1$ AFCI phase located between the QSH and SDW phases, the degeneracy of $\delta_{\mathrm{MF}}^{\sigma}$ is lifted: one spin component remains inside the shaded region of Fig.~\ref{fig_mass}(b), while the other goes outside. Consequently, the two bands split into topologically nontrivial and trivial sectors, yielding a net Chern number $C=1$. As shown in Fig.~\ref{fig_mass}(a), the phase boundaries extracted from the effective Dirac mass agree well with those obtained by the conventional Berry-curvature calculation based on the MF Hamiltonian~\eqref{eq_ham_MF}.

A key feature of $\delta_{\mathrm{MF}}^{\sigma}$ in Fig.~\ref{fig_mass}(b) is that its spin degeneracy is lifted only in the AFCI phase, while it remains intact in all other phases, including the SDW. This behavior is consistent with the MF results for AFM orders~\cite{JiangK_2018}: the degeneracy lifting in the AFCI phase arises from a symmetry-breaking AFM order along the $z$ axis [see Eq.~\eqref{eq_delta_delta}], whereas in the SDW phase the AFM correlations are confined to the $xy$ plane, preserving $z$-symmetry. This contrasts with the situation in the HH model~\cite{HeWX_2024}, and explains why $\delta_{\mathrm{MF}}^{\sigma}$ remains degenerate in the SDW phase. 

While the Hartree-Fock approach used here captures the AFCI phase qualitatively, it is prone to overestimate symmetry-breaking tendencies and may therefore yield quantitative discrepancies. More sophisticated treatments, such as slave-boson calculations that account for strong correlations more accurately, shift the phase boundaries to higher $U$ but leave the topological structure of the phase diagram intact~\cite{JiangK_2018}. This highlights the robustness of the AFCI phase and confirms its intrinsic origin in the interplay among SOC, sublattice symmetry breaking, and Hubbard interactions.

\section{Summary and conclusion}
\label{sec_summary}

In this work, we present compelling evidence for the AFCI phase in the phase diagram of the KMH model with a staggered sublattice potential. Our ED results---based on analyses of the excitation gap, anisotropic AFM correlations, fidelity susceptibility with respect to TBCs, and (spin) Chern number evaluations---demonstrate the emergence of a topologically nontrivial phase that spontaneously breaks TRS. This phase arises from the interplay between band topology and electron-electron interactions. It is characterized by reoriented AFM correlations coexisting with the breakdown of the shared topological character between the two spin species, with one being topologically nontrivial and the other trivial, giving rise to a quantized Chern number $C=1$. At the MF level, this phenomenon can be understood in terms of an effective spin-dependent sublattice potential. 

More specifically, we show that although the KMH Hamiltonian itself respects TRS, the system develops an intrinsic instability towards TRS breaking when subject to Hubbard-driven AFM perturbations. This instability emerges near the CDW-QSH phase boundaries, where the excitation gap is small and AFM correlations are anomalously enhanced. It manifests as a violation of the adiabatic continuity condition under TBCs: true level crossings occur at gap-closing points in $\bm{\theta}$ space, accompanied by singularities in the lattice field strength of the Berry curvature. To address this, we introduce a modified algorithm for Chern number calculations that tracks the smooth evolution of the target wavefunction across the level-crossing region in $\bm{\theta}$ space. This method successfully captures the TRS-breaking character of the AFCI phase and yields a robust Chern number $C=1$. 

Experiments with cold atoms in optical lattices have established highly tunable platforms for simulating a variety of important models in condensed matter physics (see, e.g., Refs.~\cite{Jordens_2008,Soltan_2011,Jotzu_2014,PanJW_2016,PanJW_2024}). With further developments, the investigation of the AFCI phase in KMH-related systems on these platforms may become accessible.

In parallel, recent advances in transition-metal dichalcogenide moir\'e superlattices---particularly in \ce{MoTe2} and \ce{WSe2}---offer another promising avenue for investigating the interplay between strong correlations and topology~\cite{MacDonald_2019,ShanJie_2021,ShanJie_2023,Cai_2023,XuXiaodong_2023,Xu_2023,Feldman_2024,Zhao_2024,XiaoDi_2024a,ShanJie_2024}. In such twisted bilayer systems, for example, the effective sublattice potential relevant to the KMH model can be tune by applying an out-of-plane electric field~\cite{XiaoDi_2024b}, thereby providing a controlled setting for examining AFCI-related phenomena.

\begin{acknowledgments}
We thank Rubem Mondaini and Eduardo V. Castro for the helpful discussions.
B.-Q.W and H.L. acknowledge support from the National Natural Science Foundation of China (NSFC; Grants No. 12174168 and No. 12247101). 
C.S. acknowledges support from NSFC (Grant No.12104229) and the Fundamental Research Funds for the Central Universities (Grant No. 30922010803).
H.-G.L. acknowledges funding from NSFC (Grants No. 11834005) and the National Key Research and Development Program of China (Grant No. 2022YFA1402704).
T.T. is partly supported by the Japan Society for the Promotion of Science, KAKENHI (Grant No. 24K00560 and No. 25H01248) from the Ministry of Education, Culture, Sports, Science, and Technology, Japan.
\end{acknowledgments}


%
\end{document}